\documentclass[tighten,twocolumn,onecolappendix]{aastex702}

\usepackage{amsmath,footmisc}
\usepackage[T1]{fontenc}

\def\d{\partial}

\def\i{\mathrm{i}}

\newcommand{\g}{\sqrt{-g}}

\newcommand{\rp}{r_{+}}

\newcommand{\OmegaF}{\Omega_{\rm F}}
\newcommand{\OmegaH}{\Omega_{\rm H}}

\newcommand{\B}{\mathcal{B}}
\newcommand{\E}{\mathcal{E}}

\newcommand{\Fd}{\vphantom{F}^\star \! F}
\newcommand{\Jd}{\vphantom{J}^\star \! J}
\newcommand{\D}{\Delta}

\newcommand{\Sigp}{\Sigma_{+}}
\newcommand{\Pp}{\mathcal{P}_{+}}
\newcommand{\Pii}{\Pi}

\newcommand{\cE}{\mathcal{E}}
\newcommand{\cB}{\mathcal{B}}
\newcommand{\cA}{\mathcal{A}}

\def\lsim{\mathrel{\raise.3ex\hbox{$<$\kern-.75em\lower1ex\hbox{$\sim$}}}}
\def\gsim{\mathrel{\raise.3ex\hbox{$>$\kern-.75em\lower1ex\hbox{$\sim$}}}}
\def\gtwid{\mathrel{\raise.3ex\hbox{$>$\kern-.75em\lower1ex\hbox{$\sim$}}}}
\def\proptwid{\mathrel{\raise.3ex\hbox{$\propto$\kern-.75em\lower1ex\hbox{$\sim$}}}}

\newcommand{\ab}[1]{\left|#1\right|}

\newcommand{\br}[1]{\left[#1\right]}

\newcommand{\pa}[1]{\left(#1\right)}
\newcommand{\ed}{\mathop{}\!\mathrm{d}}
\newcommand{\pd}{\mathop{}\!\partial}
\DeclareMathOperator\sign{sign}

\graphicspath{{figures/}}
\definecolor{indigo(dye)}{rgb}{0.0, 0.25, 0.42}

\defcitealias{PaperI}{EHTC~I}
\defcitealias{PaperV}{EHTC~V}
\defcitealias{PaperVI}{EHTC~VI}
\defcitealias{PaperVII}{EHTC~VII}
\defcitealias{PaperVIII}{EHTC~VIII}
\defcitealias{Paper2}{Paper II}
\defcitealias{Paper3}{Paper III}

\begin{document}

\title{Black Hole Polarimetry: Universal Polarization of Synchrotron Radiation at the Horizon}

\correspondingauthor{Andrew Chael}
\author[0000-0003-2966-6220]{Andrew Chael}
\email[show]{andrew.chael@nbi.ku.dk}
\affiliation{Niels Bohr International Academy, Niels Bohr Institute, Blegdamsvej 17, DK-2100 Copenhagen Ø, Denmark}

\author[0000-0002-1559-6965]{Alexandru Lupsasca}
\email{alexandru.v.lupsasca@vanderbilt.edu}
\affiliation{Department of Physics \& Astronomy, Vanderbilt University, Nashville TN 37212, USA}

\author[0000-0001-6952-2147]{George~N.~Wong}
\email{gnwong@ias.edu}
\affiliation{Princeton Gravity Initiative, Princeton University, Princeton NJ 08544, USA}
\affiliation{School of Natural Sciences, Institute for Advanced Study, Princeton NJ 08540, USA}

\author[0000-0001-8053-4392]{Zachary Gelles}
\email{zgelles@princeton.edu}
\affil{Department of Physics, Princeton University, Princeton NJ 08540, USA}
\affiliation{Princeton Gravity Initiative, Princeton University, Princeton NJ 08544, USA}

\author[0000-0001-9185-5044]{Eliot Quataert}
\email{quataert@princeton.edu}
\affiliation{Department of Astrophysical Sciences, Princeton University, Princeton NJ 08544, USA}
\affiliation{Princeton Gravity Initiative, Princeton University, Princeton NJ 08544, USA}

\begin{abstract}
Polarized images of a black hole encode the direction of electromagnetic energy flow near its event horizon.
Measuring polarization from near-horizon emission can help determine whether this energy flow is powered by the accreting plasma or the black hole spin.
Here we consider the linear polarization of synchrotron radiation emitted from the base of horizon-threading field lines in a time-stationary, axisymmetric, and degenerate Kerr magnetosphere with nonzero poloidal current.
We show that the observed polarization pattern displays universal behavior: it is completely determined by the black hole spin and observer inclination and is independent of the magnetic field geometry.
We derive a simple analytic formula for this spin-dependent horizon polarization pattern.
We find that this predicted pattern is also approached in time-averaged images from General Relativistic Magnetohydrodynamic simulations.
Future observations with Very-Long-Baseline Interferometry at microarcsecond resolution could detect the trend of polarization toward the unique horizon value in M87*.
Such observations may enable new measurements of black hole spin and provide evidence that magnetic field lines thread the horizon, a necessary condition for spin-energy extraction via the Blandford--Znajek process.
\end{abstract}

% Ensure references to EHT papers appear in chronological order
%\nocite{PaperI}
\nocite{PaperIV}
\nocite{PaperV}
\nocite{PaperVI}
\nocite{PaperVII}
\nocite{PaperVIII}
%\nocite{PaperI_SgrA}
\nocite{PaperIX}
\nocite{PaperVII_SgrA}
\nocite{PaperVIII_SgrA}
\nocite{EHT_2021}

\defcitealias{BHPI}{BHPI}
\defcitealias{BHPII}{BHPII}

\section{Introduction}
The near-horizon environments of most accreting supermassive black holes in the Universe are characterized by hot, magnetized plasmas that produce most of their bolometric luminosity from synchrotron radiation \citep{Ho_2008,Yuan_Narayan_2014}.
The polarization of this synchrotron radiation is uniquely sensitive to the coupled effects of the black hole's strong gravity, its magnetic field, and the surrounding plasma's composition and distribution function.

The Event Horizon Telescope has directly probed the near-horizon magnetospheres of the black holes M87* \citep{PaperVII,PaperIX} and Sgr\,A* \citep{PaperVII_SgrA} with Very-Long-Baseline Interferometry (VLBI) observations at 230 GHz.
Images of both Sgr\,A* and M87* over multiple epochs feature similar organized spiral patterns of electric vector position angles (EVPA) around the central emission ring.
These images have provided substantially more information about the accretion flows around both black holes than total intensity images alone.
In particular, comparing the EHT's polarized images of both sources to images from General Relativistic Magnetohydrodynamic (GRMHD) simulations \citep{PaperVIII,PaperVIII_SgrA} has shown that both black holes are likely surrounded by strongly magnetized or magnetically arrested \citep[MAD;][]{Narayan2003} accretion flows, where ordered poloidal magnetic fields strongly affect the dynamics of both plasma inflow and outflow.

Ordered poloidal magnetic fields threading the event horizon are a necessary ingredient for extracting the spin energy of a black hole in the \citet{BlandfordZnajek1977} (BZ) process.
The BZ process is ubiquitous in GRMHD simulations of magnetized accretion flows and is widely thought to be the most likely power source of extragalactic jets throughout the Universe, including the paradigmatic example of M87*'s multi-kiloparsec jet.
Despite the ubiquity of extragalactic jets in the Universe and their importance for understanding the co-evolution of supermassive black holes and their host galaxies, there is no conclusive demonstration that BZ energy extraction powers jets.

In earlier papers in this series, we showed that the structure of polarized images of black holes directly probes the direction of electromagnetic energy flow near their event horizons \citep[][hereafter \citetalias{BHPI}]{BHPI}.
In particular, the rotation sense of the polarization spiral encoded in the phase of the second azimuthal Fourier mode $\angle\beta_2$ of the linear polarization \citep{Palumbo2020} directly depends on whether electromagnetic energy flows inward or outward from the black hole, assuming the sign of the field line angular velocity $\OmegaF$ is known.
This diagnostic is robust in both MAD GRMHD simulations and analytic GRMHD inflow models \citep[][hereafter \citetalias{BHPII}]{BHPII}.
Furthermore, as first noted by \citet{Palumbo2020} in GRMHD simulations, the polarization pattern encoded in $\angle\beta_2$ is sensitive to black hole spin.
We explained this trend by connecting the observed $\angle\beta_2$ to the spin-dependent windup of magnetic fields in a BZ outflow, both in its average value \citepalias{BHPI} and in its radial evolution across the image plane \citepalias{BHPII}.

In both Sgr\,A* and M87*, polarized EHT images \citep{PaperVII, PaperVII_SgrA, EHT_2021} currently probe radii $r\approx 5\,r_{\rm g}$, where the gravitational radius is $r_{\rm g}=GM/c^2$ and $M$ is the black hole mass.
To directly probe the predicted energy extraction from black holes and test the BZ mechanism as the source of extragalactic jet power, we must track the energy outflow on field lines ever closer toward the event horizon by measuring polarization from dimmer, increasingly redshifted emission interior to the primary emission ring toward the projected image of the horizon itself \citep[the ``inner shadow'';][]{Chael2021}.

Away from the horizon, the observed synchrotron polarization depends on both the local magnetic field configuration and the gravitational parallel transport of the polarization vector; the reversal of the polarization direction from strong parallel transport in increasingly lensed sub-images has been noted as a strong diagnostic of black hole spin \citep{Himwich2020}.
In this paper, we show that the observed synchrotron polarization from emitters asymptotically approaching the event horizon takes on a \emph{universal} form when the magnetosphere is time-stationary, axisymmetric, and infinitely conductive with nonzero poloidal current; it is independent of the specific configuration of the magnetic field and depends only on the black hole spin $a_*$ and observer inclination $\theta_{\rm o}$.

This universality was alluded to (but not shown) in \citetalias{BHPI} with reference to a forthcoming paper; it was first demonstrated for direct ($n=0$) emission from the equatorial plane by \citet{Hou2024}.
Here, we present a unified argument that extends the result to all source colatitudes $\theta_{\rm s}$ on the horizon and to all image orders $n$.
Our result can be loosely understood as a consequence of the increasingly toroidal magnetic field and increasingly radial emitter velocity close to the horizon picking out the polar direction $\hat{\theta}$ as the dominant direction of the polarization in the emitter frame at the horizon.
In a certain sense, the horizon polarization maps the parallel-transported lines of longitude across the multiply-lensed images of the black hole horizon.

We first review the propagation of polarized light in the Kerr spacetime in Section~\ref{sec:poltransport} before specializing to its uniquely simple form for synchrotron radiation from degenerate magnetospheres in Section~\ref{sec:syncpol}.
In Section~\ref{sec:horizonsig}, we derive the key result: the observed synchrotron EVPA from a degenerate, stationary, axisymmetric magnetosphere becomes independent of the electromagnetic field as the emitters approach the horizon and takes on a strikingly simple form (Equation~\eqref{eq:univevpa}).
We then validate our analytic formula for near-horizon polarization against numerical images of thin shells of near-horizon plasma from the general relativistic radiative transfer code \texttt{ipole} \citep{Moscibrodzka2018}.

In Section~\ref{sec:detection}, we explore the prospects for detecting the near-horizon polarization signature with future VLBI observations of M87*.
Surprisingly, we show that the universal horizon value is approached in time-averaged images of three-dimensional GRMHD simulations when the emission region is sufficiently compact.
The proposed space-VLBI Black Hole Explorer mission \citep[BHEX; ][]{BHEX} would extend millimeter VLBI baselines to Earth orbit and achieve resolutions of a few $\mu$as.
We show that this enhanced resolution could in principle resolve the characteristic swing in $\angle\beta_2$ toward the universal horizon value in the equatorial emission region of M87*.
We discuss the implications of our result in Section~\ref{sec:discussion} and conclude in Section~\ref{sec:conclusions}.

\section{Polarization Transport in Kerr}
\label{sec:poltransport}

This section reviews the description of polarization of light and its parallel transport in the Kerr spacetime, which we will need in subsequent sections.
All results presented in this section are known, but we collect them here in a simple, unified presentation.
We show that the simplicity of parallel transport in the Kerr spacetime ultimately follows from its hidden symmetry, encoded in the Killing-Yano tensor. Throughout, we use units normalized such that $c=G=1$.

\subsection{Kerr Metric}
A Kerr black hole is fully described by its mass $M$ and angular momentum $J=aM$, which is limited to the range $0\leq a\leq M$; we define $a_*=a/M$ to be the dimensionless spin that obeys $0\leq a_*\leq 1$.
In Boyer-Lindquist coordinates $(t,r,\theta,\phi)$, the Kerr metric
\begin{gather}
    \label{eq:Kerr}
	ds^2=-\frac{\Delta}{\Sigma}\pa{\ed t-a\sin^2{\theta}\ed\phi}^2+\frac{\Sigma}{\Delta}\ed r^2+\Sigma\ed\theta^2\notag\\
	+\frac{\sin^2{\theta}}{\Sigma}\br{\pa{r^2+a^2}\ed\phi-a\ed t}^2,
\end{gather}
is expressed in terms of three poloidal functions
\begin{gather}
    \Delta = r^2 + a^2 - 2Mr, \qquad \Sigma = r^2 + a^2\cos^2\theta, \nonumber \\
    \Pi = (r^2+a^2)^2 - a^2\Delta \sin^2\theta.
\end{gather}
The Kerr event horizon occurs when $\Delta=0$ at radius
\begin{equation}
    r_+ = M + \sqrt{M^2 - a^2}.
\end{equation}
Normal observers in the Boyer-Lindquist Zero Angular Momentum Observer (ZAMO) frame $\eta^\mu$ have angular velocity $\omega = \eta^\phi/\eta^t = 2aMr/\Pi$. At the event horizon, the angular velocity of the normal observer matches the horizon angular velocity: \begin{equation}
    \OmegaH = \frac{a}{2Mr_+} = \omega |_{r=r_+}.
\end{equation}
Finally, we note that the metric determinant is $\sqrt{-g}=\Sigma \sin\theta$, and the Levi-Civita tensor is defined as ${\epsilon^{\mu\nu\alpha\beta}=-\left[\mu\nu\alpha\beta\right]/\sqrt{-g}}$, where $\left[\mu\nu\alpha\beta\right]$ is the completely antisymmetric symbol.

\subsection{Null Geodesics in Kerr}
We consider an observer lying at a large distance $r_{\rm o}\gg M$ from a Kerr black hole, at a polar inclination $\theta_{\rm o}$ from its axis of rotation.
The image plane of such an observer perpendicular to the ``line-of-sight'' to the black hole is covered by \citet{Bardeen1973} coordinates $(\alpha,\beta)$.\footnote{The $+\hat{\alpha}$ vector is aligned with the $+\hat{\phi}$ azimuthal direction at infinity, and the $+\hat{\beta}$ axis is aligned with the $-\hat{\theta}$ direction, along the projected black hole spin vector.}
Null geodesics arriving on the image plane at a given $(\alpha,\beta)$ coordinate have conserved specific angular momentum $\lambda$ and Carter constant $\eta$:
\begin{subequations}
\label{eq:lambdaeta}
\begin{align}
    \lambda &= -\alpha\sin\theta_{\rm o}, \label{eq:lambda} \\
    \eta &= (\alpha^2-a^2)\cos^2\theta_{\rm o} + \beta^2.
    \label{eq:eta}
\end{align}
\end{subequations}
These conserved quantities ($\lambda$, $\eta$) determine the photon's four-momentum, and hence its trajectory, everywhere:\footnote{The signs $\pm_r$ and $\pm_\theta$ in Equation~\eqref{eq:PhotonMomentum} switch at radial and angular turning points where $p^r=0$ and $p^\theta=0$, respectively. For outgoing geodesics arriving at the observer screen, $\pm_{\theta_\mathrm{o}}=\mathrm{sign}(\beta)$.}
\begin{align}
	\label{eq:PhotonMomentum}
	\frac{p_\mu\ed x^\mu}{E}=-\ed t\pm_r\frac{\sqrt{\mathcal{R}(r)}}{\Delta(r)}\ed r\pm_\theta\sqrt{\Theta(\theta)}\ed\theta+\lambda\ed\phi.
\end{align}
where $E=-p_t$ is the photon energy, and $\mathcal{R}(r)$ and $\Theta(\theta)$ are radial and angular potentials:
\begin{subequations}
\begin{align}
    \mathcal{R}(r) &= (r^2 + a^2 - a \lambda)^2 - \Delta\left[\eta + (\lambda-a)^2\right], \label{eq:Rpot}\\
    \Theta(\theta) &= \eta + a^2\cos^2\theta - \lambda^2/\tan^2\theta. \label{eq:Tpot}
\end{align}
\end{subequations}
We will find it useful later to introduce a shifted coordinate on the image plane $\nu$:
\begin{equation}
\label{eq:nu}
	\nu=-\alpha-a\sin{\theta_{\rm o}} = \lambda\csc\theta_{\rm o} -a\sin{\theta_{\rm o}}.
\end{equation}
Furthermore, we note that the $\beta$ coordinate of a photon on the observer screen is simply the square root of its angular potential at infinity:
\begin{equation}
\label{eq:betawrtTheta}
\beta = \pm_\theta\sqrt{\Theta(\theta_{\rm o})}.
\end{equation}

The trajectory $x^\mu(\tau)$ of a photon arriving at a screen location $(\alpha,\beta)$ can be computed fully analytically as a function of its affine parameter $\tau$ \citep[e.g.][]{GrallaLupsasca2020,GrallaLupsasca2020b}. Critically, photon trajectories that arrive at the observer screen can be divided into two classes: those that experience a radial turning point where $p^r=0$ and those that do not. Null geodesics that experience a radial turning point extend backwards to null infinity, while geodesics without a radial turning point cross the event horizon. On the $(\alpha,\beta)$ image plane, these two classes of geodesics are separated by a convex \emph{critical curve} enclosing the origin, which represents the asymptotic image of bound photon orbits around the black hole \citep[e.g.][]{JohnsonLupsasca2020}.\footnote{The shape of the critical curve can be determined by finding the ($\alpha$,$\beta$) points corresponding to the ($\lambda$,$\eta$) trajectories of bound photon orbits by inverting Equation~\eqref{eq:lambda}--\eqref{eq:eta}. Alternatively, one can map out whether or not a ray from a given ($\alpha$,$\beta$) has a radial turning point by considering the roots of the radial function $\mathcal{R}(r)$ \citep{GrallaLupsasca2020b}.}

Null geodesics that arrive at the image plane inside the critical curve can thus be traced back to the event horizon. Emission directly from the horizon at $r=r_+$ will experience infinite redshift and thus not be directly observable; nevertheless, rays arriving inside the critical curve can be thought to provide an \emph{asymptotic image} of emission approaching the horizon in the limit $r\rightarrow r_+$.
\begin{figure*}
\centering
\includegraphics[width=\textwidth]{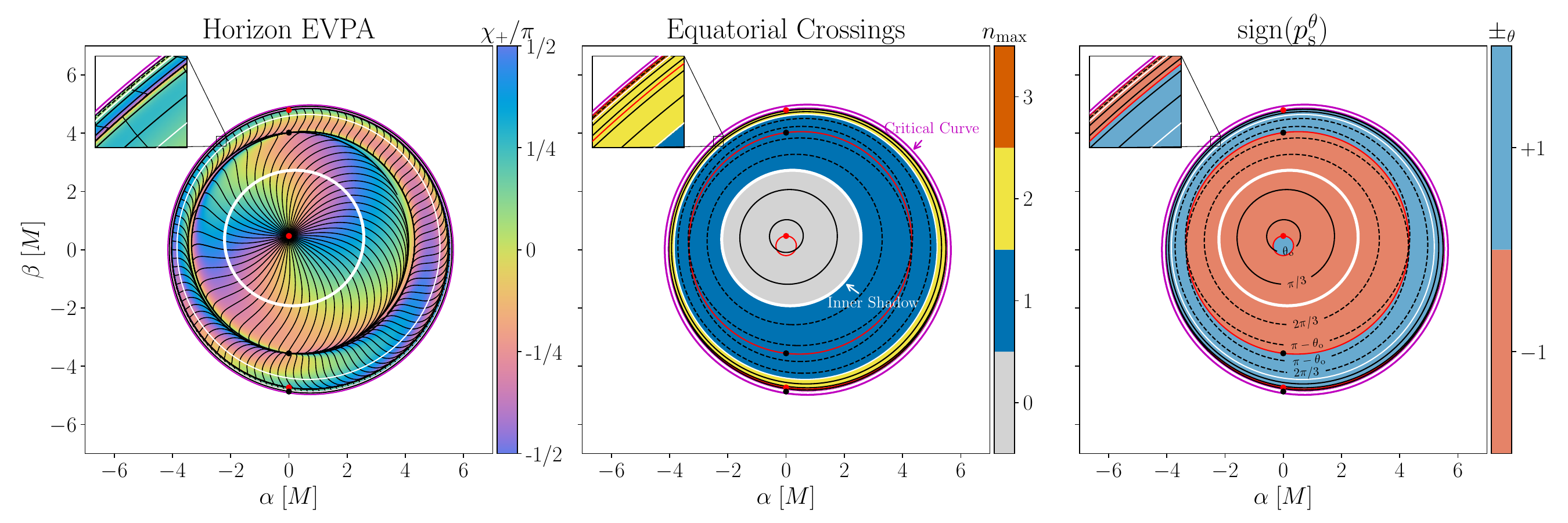}
\caption{
The horizon EVPA, equatorial lensing structure, and sign of the outgoing polar momentum at the horizon for a Kerr black hole with spin $a_* = 0.9375$ viewed at inclination $\theta_{\rm o} = 20\deg$.
\emph{Left:} The horizon EVPA $\chi_+$ plotted in a cyclic colormap; integrated streamlines following the EVPA are overplotted in black.
\emph{Center:} The number of equatorial crossings $n$ made by null geodesics arriving at the observer screen.
Contours of constant polar angle $\theta_{\rm s}$ on the horizon are shown in solid black lines for the northern hemisphere ($\theta_{\rm s}<\pi/2$) and dashed black lines for the southern hemisphere ($\theta_{\rm s}>\pi/2$). For $\theta_{\rm s}<\pi-\theta_{\rm o}$ these contours are closed circles; for $\theta_{\rm s}>\pi-\theta_{\rm o}$ they become banana-like shapes enclosing the image of the poles.
\emph{Right:} The sign of the polar momentum of the outgoing null geodesic $\sign(p^\theta_{\rm s})$ at the horizon, which enters into the equation for the horizon EVPA (Equation~\eqref{eq:univevpa}).
Red contours indicate boundaries where the sign of the outgoing polar momentum at the horizon switches when $\Theta(\theta_{\mathrm{s}})=0$.
\emph{All panels:} The outer magenta contours indicate the critical curve, and white contours indicate successive images of the intersection of the event horizon with the equatorial plane. Successive images of the black hole's north and south poles are indicated by black and red dots.
Insets in the top left show a zoomed-in view of the $n=1$ and $n=2$ photon ring structure.
}
\label{fig:explainer}
\end{figure*}

\subsection{Polarization of Light and its Propagation}

A photon propagating in Kerr may carry linear polarization, described by a spacelike vector $f^\mu$. The polarization vector is orthogonal to the direction of propagation and is parallel-transported along the ray:
\begin{gather}
\label{eq:PolarizationConditions}
    f\cdot p=0, \qquad p^\mu\nabla_\mu f^\nu=0.
\end{gather}
Given any such polarization vector $f^\mu$, any other vector differing by an arbitrary shift $\zeta$ along the photon momentum,
\begin{align}
	\label{eq:Shift}
	\tilde{f}^\mu=f^\mu+\zeta p^\mu,
\end{align}
also obeys the polarization conditions (Equation~\eqref{eq:PolarizationConditions}).
Physically, all the vectors related by Equation~\eqref{eq:Shift} describe the same polarization state.

As explained in Appendix~\ref{app:KerrSymmetry}, the hidden symmetries of the Kerr metric enable photons propagating in this spacetime to carry an additional conserved scalar quantity besides the energy $E$, angular momentum $\lambda$ and Carter constant $\eta$.
The Penrose-Walker constant $\kappa$ \citep{Walker_Penrose_1970} is:
\begin{equation}
\label{eq:KYPW}
\kappa=\br{\star J_{\mu\nu}+iJ_{\mu\nu}}p^\mu f^\nu.
\end{equation}
In Equation~\eqref{eq:KYPW}, $J_{\mu\nu}$ is the antisymmetric \emph{Killing-Yano tensor}
(Equation~\eqref{eq:KYBL}) and $\star J_{\mu\nu}$ is its dual (Equation~\eqref{eq:CKYBL}). The symmetries of $J_{\mu\nu}$ imply the conservation of the complex constant $\kappa$ for any vector $f^\mu$ parallel transported along $p^\mu$ (see \autoref{app:KerrSymmetry}). Since the Killing-Yano tensor $J_{\mu\nu}$ is antisymmetric, by Equation~\eqref{eq:KYPW}, $\kappa$ is invariant to any shifts of $f^\mu$ along $p^\mu$.

The Penrose-Walker constant is often written as
\begin{equation}
	\label{eq:ABdef}
	\kappa=\kappa_1+i\kappa_2 =\pa{\mathcal{A}-i\mathcal{B}}\pa{r-ia\cos{\theta}},
\end{equation}
where $\mathcal{A}$ and $\mathcal{B}$ are real scalars that are naturally defined in terms of a local Newman-Penrose null basis (see Appendix~\ref{app:NPdefs}).

The norm of the Penrose-Walker constant encodes the magnitude of the polarization vector,\footnote{The norm of the PW constant is $\ab{\kappa}^2=\br{\eta+(a-\lambda)^2}f^2=K_{\mu\nu}p^\mu p^\nu \, f^2/E^2$, where $K_{\mu\nu}$ is the Kerr Killing tensor.}
while information about the polarization direction is stored in the phase of $\kappa$, or equivalently in the ratio $z\equiv\kappa_1/\kappa_2$.

The existence and conservation of the Penrose-Walker constant
effectively solves the parallel transport problem for the linear polarization of light propagating in Kerr in the absence of local emission, absorption, or Faraday rotation.
Given a photon emitted with some source four-momentum $p_{\rm s}^\mu$ and polarization $f_{\rm s}^\mu$, one can compute the conserved quantity $\kappa$ at the source; later, when the photon is observed with four-momentum $p_{\rm o}^\mu$, one can immediately reconstruct its local polarization $f_{\rm o}^\mu$ by inverting Equation~\eqref{eq:ABnp} at the observer while imposing orthogonality $f\cdot p=0$ and manually fixing the gauge (Equation~\eqref{eq:Shift}).

\subsection{Polarization on the Image Plane}
We now turn to the description of the polarization vector on the image plane for an observer at infinity (see Appendix~A of \citealt{Himwich2020} for a full derivation).
In our convention, the EVPA $\chi$ is the angle of the plane of linear polarization at the observer screen counterclockwise from the $+\hat{\beta}$ axis.
In this convention, the observed EVPA at infinity can be computed from the phase of the Penrose-Walker constant by:\footnote{In Equation~\eqref{eq:chiexplicit2} and other equations where $\chi$ is given as the sum or difference of arctangents, it should be assumed that we take the modulus $\chi \rightarrow (\chi-\pi/2 \mod \pi) -\pi/2$ so that $\chi\in (-\pi/2,\pi/2)$.}
\begin{subequations}
\label{eq:chiexplicit}
\begin{align}
    \label{eq:chiexplicit1}
     \chi &= \arctan\left(\frac{\nu\kappa_1-\beta \kappa_2}{\beta\kappa_1+\nu \kappa_2}\right), \\
     \label{eq:chiexplicit2}
      &= \arctan\left(z\right) - \arctan\left(\frac{\beta}{\nu}\right),
\end{align}
\end{subequations}
where we define the ratios
\begin{equation}
\label{eq:zdef}
z \equiv \frac{\kappa_1}{\kappa_2} = \frac{a\cos\theta+r\mathcal{Z}}{r-a\cos\theta\mathcal{Z}},
\end{equation}
and
\begin{equation}
\label{eq:Rdef}
\mathcal{Z} \equiv -\frac{\mathcal{A}}{\mathcal{B}}.
\end{equation}

Thus, if we can determine the ratio $\mathcal{Z} = -\mathcal{A}/\mathcal{B}$ at the source for a polarized emitter in the Kerr spacetime, we completely determine the observed EVPA from that source at infinity.
In astrophysical settings, we typically have to deal with the more complicated situation of integrating the radiative transfer equation along each null geodesic that arrives on the image plane (accounting for continuous emission, absorption, Faraday rotation, etc.). However, one can show that the rotation of the local emitting frame due to parallel transport at each point along the ray only depends on the point-to-point $\chi$ defined in Equation~\eqref{eq:chiexplicit} \citep[see][equations 47-48]{Dexter2016}.

Finally, we note that the EVPA structure in ring-like black hole images often exhibits approximate azimuthal symmetry on the image plane. In this case, it is useful to summarize the EVPA structure with the $\beta_2$ statistic \citep{Palumbo2020}, defined as the coefficient of the $m=2$ angular mode of the complex polarization $|P|e^{2i\chi}$:
\begin{equation}
\label{eq:beta2}
\beta_2(b) = \frac{\int |P|e^{2i\chi}e^{-2i\varphi} b \, d\varphi}{\int I b \, d\varphi},
\end{equation}
where $(b,\varphi)$ are polar coordinates on the image plane and $|P|$ and $I$ are the polarized and the total specific intensities.
When the EVPA structure is azimuthally symmetric, $\chi(\varphi) =\chi_0+\varphi$, the phase of $\beta_2$ is simply equal to $\angle\beta_2=2\chi_0$, where $\chi_0$ is the EVPA on the $+\hat\beta$ axis at angle $\varphi=0$.

\begin{figure*}
\centering
\includegraphics[width=\textwidth]{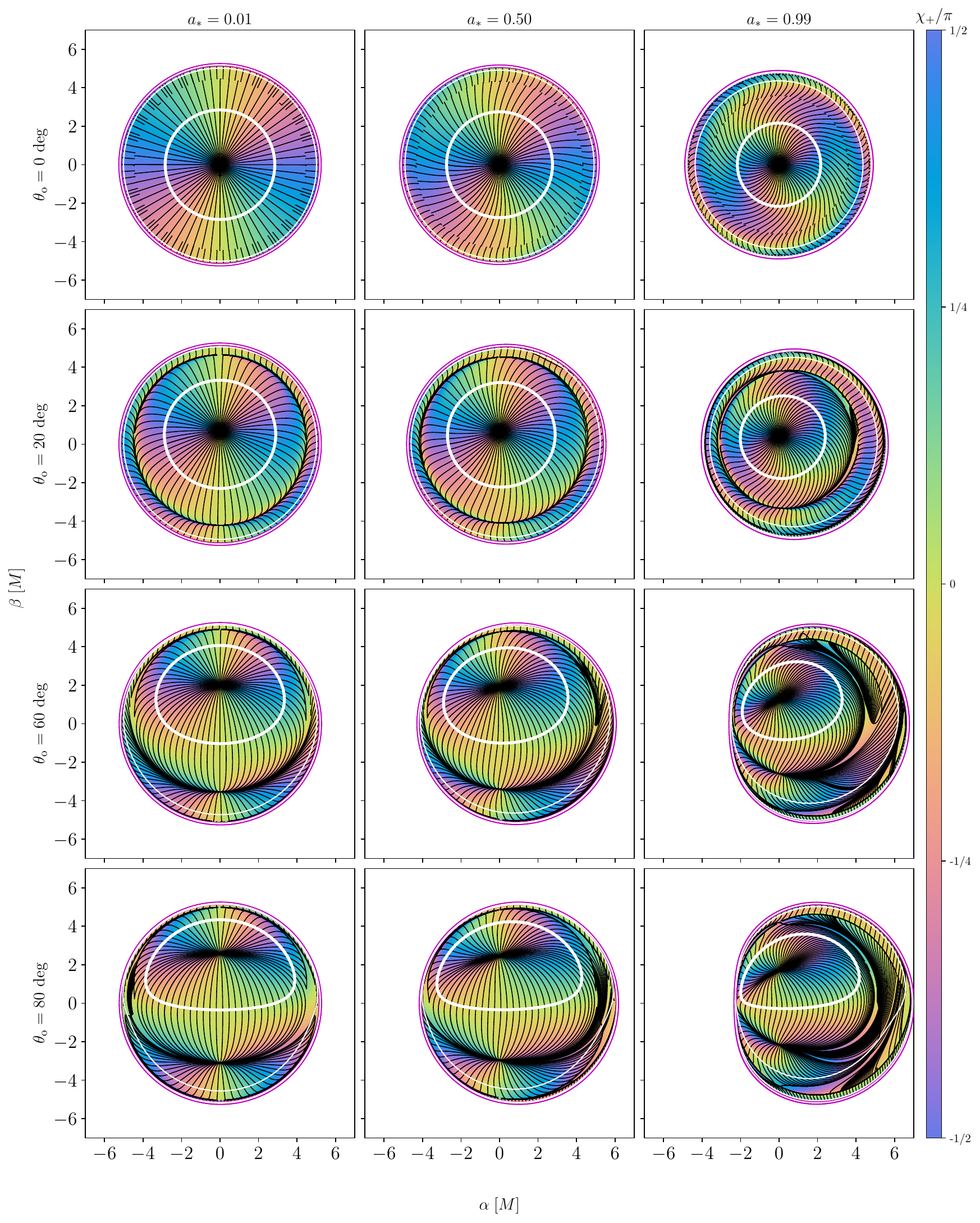}
\caption{The universal horizon formula (Equation~\eqref{eq:univevpa}) plotted inside the critical curve for black holes viewed at inclinations $\theta_{\rm o}=0,20,60, 80\deg$ (rows) and spins $a_*=0.01,0.5,0.99$ (columns). In all panels, white contours indicate successive images of the black hole's equator $\theta_{\rm s}=\pi/2$, and integrated streamlines following the local EVPA are plotted in black.}
\label{fig:examplehorizons}
\end{figure*}

\section{Synchrotron Polarization from Degenerate Electromagnetic Fields}
\label{sec:syncpol}
In this section, we turn to the specific properties of synchrotron polarization in the Kerr spacetime before specializing to synchrotron emission from \emph{degenerate} electromagnetic fields, such as those that obey the equations of GRMHD.
\subsection{Synchrotron Polarization}
The linear polarization of synchrotron radiation is emitted perpendicular to the local magnetic field $b$ and the emitter's four-velocity $u$ \citep[e.g.][]{RybickiLightman, Dexter2016}. The polarization vector $f_{\rm s}$ describing synchrotron radiation at its point of emission must therefore satisfy three relations:\footnote{
We can arrive at the conditions in Equation~\eqref{eq:Conditions} by requiring that $f^{(\hat a)}$ be purely spatial and perpendicular to the photon momentum and magnetic field in a locally orthonormal frame moving with velocity $u^\mu$.}
\begin{align}
	\label{eq:Conditions}
	f_{\rm s}\cdot p_{\rm s}=0,\qquad
	f_{\rm s}\cdot b_{\rm s}=0,\qquad
	f_{\rm s}\cdot u_{\rm s}=0.
\end{align}
These three conditions fix three of the four components of the source polarization vector $f_{\rm s}^\mu$, leaving only its length undetermined. This length is determined by the detailed emission physics of the source (e.g. the polarized synchrotron emissivity for a thermal plasma at a given temperature, magnetic field strength, and frequency, \citealt{Pandya2016}).
The conditions in Equation~\eqref{eq:Conditions} are automatically satisfied by an explicit construction:
\begin{equation}
\label{eq:polvec}
    f_{\rm s}^\mu = -\epsilon^{\mu\nu\kappa\lambda}\, p_{\nu} \, b_\kappa u_\lambda,
\end{equation}
where we drop explicit source labels for $p^\mu_{\rm s}, u^\mu_{\rm s}$ and $b^\mu_{\rm s}$.\footnote{Note that $f^\mu_{\rm s}$ given by Equation~\eqref{eq:polvec} will generally not be in temporal gauge, $f^t_{\rm s}\neq 0$.}
Note that Equation~\eqref{eq:polvec} shows that any velocity parallel to the local magnetic field does not affect the synchrotron polarization direction.\footnote{A useful relation that emerges from Equation~\eqref{eq:polvec} is that the conserved magnitude of the polarization vector is set in the source frame by $f_{\rm s}^2 = b^2 E^2 \sin^2\theta_{\rm B} /g^2$, where $g=-E/p^\mu u_\mu$ is the redshift factor and $\theta_{\rm B}$ is the source-frame angle between the magnetic field and the photon momentum \citep{Dexter2016}.}

\subsection{Polarization from Degenerate Fields}
For a given electromagnetic field tensor $F_{\mu\nu}$, the local electric and magnetic field vectors in a timelike frame moving with velocity $u^\mu$ are:
\begin{gather}
\label{eq:emubmu}
e^\mu = u_\nu F^{\mu\nu}, \qquad b^\mu = -u_\nu \Fd^{\mu\nu}.
\end{gather}
Electromagnetic fields are said to be \emph{degenerate} and \emph{magnetically dominated} when there exists everywhere a timelike frame $u^\mu$ with zero local electric field ($e^\mu=0$) and spacelike magnetic field ($b^2>0$). In particular, these conditions are the fundamental assumptions in deriving the equations of ideal GRMHD and General Relativistic Force-Free Electrodynamics (GRFFE), both of which assume $e^\mu=0$ as a consequence of the assumed infinite conductivity of the plasma carrying the electromagnetic fields.

For degenerate, magnetically dominated fields, the Maxwell and Faraday tensors can be expressed simply in terms of the velocity $u^\mu$ and magnetic field $b^\mu$ in the frame with zero electric field \citep[e.g.][]{Gammie2003}:
\begin{subequations}
\begin{align}
    \Fd^{\mu\nu} = b^\mu u^\nu - u^\mu b^\nu, \label{eq:MaxDeg} \\
    F^{\mu\nu} = -\epsilon^{\mu\nu\kappa\lambda} b_\kappa u_\lambda. \label{eq:FarDeg}
\end{align}
\end{subequations}
A key simplification for computing the polarization direction from synchrotron-emitting particles in degenerate electromagnetic fields comes from combining Equation~\eqref{eq:polvec} and Equation~\eqref{eq:FarDeg}:
\begin{align}
\label{eq:polvec2}
    f_{\rm s}^\mu &= F^{\mu\nu}p_\nu.
\end{align}
Thus, in magnetized plasmas that obey the equations of ideal GRMHD or ideal GRFFE, the direction of the emitted polarization can simply be computed by contracting the local field strength tensor with the photon momentum.

\section{Universal Synchrotron Polarization From the Horizon}
\label{sec:horizonsig}
We now turn to the specific case of synchrotron radiation from degenerate electromagnetic fields in Kerr that are also time-stationary and axisymmetric and derive the key result of this paper: the observed synchrotron radiation from emitters approaching the horizon in these fields produces an EVPA pattern inside the critical curve that is entirely determined by the black hole spin $a$ and observer inclination $\theta_{\rm o}$.

\subsection{Stationary, Axisymmetric, Degenerate Fields}
Stationary, axisymmetric, and degenerate electromagnetic fields in Kerr have only three degrees of freedom: the poloidal flux function $\psi(r,\theta)$, the angular velocity of field lines $\OmegaF(\psi)$ and the poloidal current $I(\psi)$.
These degrees of freedom must satisfy the \citet{Znajek1977} regularity condition at the future event horizon:
\begin{equation}
    I(r_+,\theta)=\frac{4\pi M r_+ \sin\theta}{\Sigp} \left(\OmegaF-\OmegaH\right) (\partial_\theta\psi),
    \label{eq:znajek1}
\end{equation}
where $\Sigp=\Sigma(r=\rp)$. The Boyer-Lindquist lab frame magnetic field components $B^i=\star F^{it}$ are given by the flux function and current as
$B^r=\partial_\theta\psi/\sqrt{-g}$, $B^\theta=-\partial_r\psi/\sqrt{-g}$, and $B^\phi=I/\left(2\pi\Delta\sin^2{\theta}\right)$. Thus, the toroidal field $B^\phi$ diverges as $\D^{-1}$ while both $B^r$ and $B^\theta$ remain finite.

The key property of stationary, axisymmetric, degenerate fields we use in this Section is that their toroidal electric field identically vanishes: $E^\phi=-F_{t\phi}/(\D\sin^2\theta)=0$. This condition immediately follows from stationarity and axisymmetry, since by the definition of $F_{\mu\nu}$ from the vector potential $A_\mu$:
\begin{equation}
\label{eq:Ftphi0}
F_{t\phi}=\partial_t A_\phi - \partial_\phi A_t=0.
\end{equation}
We will show below that a vanishing toroidal electric field at the horizon is the key requirement for universality when determining the limiting horizon value of the observed synchrotron polarization from degenerate fields.
\begin{figure*}[t]
\centering
\includegraphics[width=\textwidth]{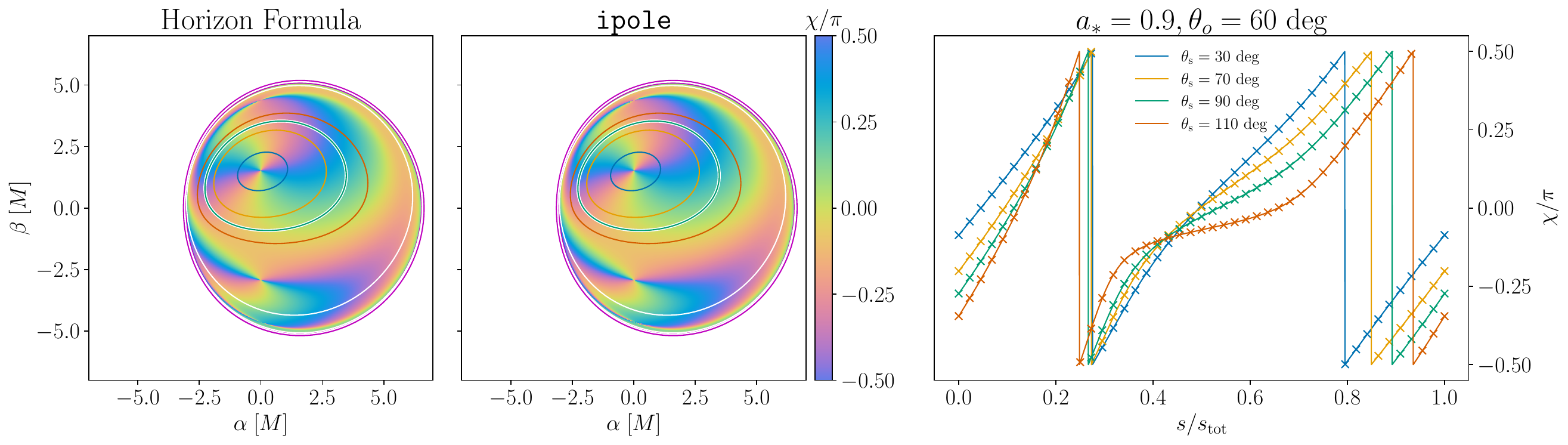}\\
\caption{Comparison of the universal horizon formula for the EVPA (Equation~\eqref{eq:univevpa}) with raytracing results from the \texttt{ipole} code for an $a_*=0.9$ black hole viewed at $\theta_{\rm o}=60\deg$.
\emph{Left:} The cyclic colormap shows the universal EVPA formula (Equation~\eqref{eq:univevpa}) inside the critical curve.
\emph{Center:} Numerical results from \texttt{ipole} obtained from confining emission to a thin shell surrounding the horizon in a monopole magnetosphere.
\emph{Right:} The EVPA $\chi$ around the $n=0$ contours of constant horizon colatitude $\theta_{\rm s}=30,70,90,110\deg$ as a function of the fractional arc length $s$ counterclockwise around the curve from the $+\hat{\beta}$-axis.
These contours are indicated on the left and center plots in the same color. The analytic horizon formula is plotted as the solid lines and EVPA values sampled from the \texttt{ipole} image are indicated with crosses.}
\label{fig:examplenumana}
\end{figure*}

\subsection{Horizon Polarization}
\label{sec:hpol}
In Appendix~\ref{app:npderivation} we calculate the Penrose-Walker constant for synchrotron polarization vectors from degenerate fields using the Newman-Penrose null basis where the Killing-Yano tensor is block diagonal. The part of the photon momentum $p^\mu$ along the outgoing principal null vector $\ell^\mu$ diverges at the horizon as $\D^{-1}$, reflecting the physical fact that outgoing rays from sources exactly at the future horizon can never reach infinity.
The Penrose-Walker constant $\kappa$ (Equation~\eqref{eq:KYPW}) also diverges at the horizon, $\kappa \propto \D^{-1}$.
However, the \emph{ratio} of the Penrose-Walker constant's real and imaginary components $z\equiv\kappa_1/\kappa_2$ and the reduced ratio $\mathcal{Z}\equiv-\mathcal{A}/\mathcal{B}$ both remain finite. They can be expanded at the horizon as
\begin{align}
z &= z_0 + \Delta z_1 + \ldots \; , \\
\mathcal{Z} &= \mathcal{Z}_0 + \Delta\mathcal{Z}_1 + \ldots \;.
\end{align}
Remarkably, in Appendix~\ref{app:npderivation} we find that for synchrotron emission from general degenerate electromagnetic fields, the asymptotic $\mathcal{Z}_0$ takes a very simple form in terms of $\beta_{\rm s}$ and $\nu_{\rm s}$, which are the analogues of the shifted Bardeen coordinates (Equations~\eqref{eq:nu} and~\eqref{eq:betawrtTheta}) evaluated at the polar angle $\theta_{\rm s}$ of the source on the horizon instead of at the observer at infinity:\footnote{Note that in Equation~\eqref{eq:betas}, the sign $\pm_{\theta_{\rm s}} = \sign\left(p^\theta_{\rm s}\right)$ will vary across the image of the event horizon, switching signs wherever $\Theta(\theta_{\rm s})=0$. This sign is not generally constant within the images of order $n$ defined by the number of equatorial crossings.}
\begin{subequations}
\label{eq:betasnus}
\begin{align}
\label{eq:betas}
\beta_{\rm s}&=\pm_{\theta_{\rm s}}\sqrt{\Theta(\theta_{\rm s})}, \\
\label{eq:nus}
\nu_{\rm s}&=\lambda\csc\theta_{\rm s}-a\sin\theta_{\rm s}.
\end{align}
\end{subequations}
The horizon value $\mathcal{Z}_0$ is completely specified by $\beta_{\rm s}$, $\nu_{\rm s}$, and the ratio of the toroidal electric and magnetic fields $q=(F_{t\phi} / \Fd_{t\phi})|_{\rp}$ at the emission point:
\begin{equation}
\label{eq:Z0general}
   \mathcal{Z}_0 = \frac{\beta_{\rm s}+q\nu_{\rm s}}{\nu_{\rm s}-q\beta_{\rm s}}.
\end{equation}
Whenever $q=0$, as is the case when the toroidal magnetic field $\star F_{t\phi}$ is nonzero\footnote{In the edge case where the field is stationary and axisymmetric ($F_{t\phi}=0$) but where the toroidal magnetic field vanishes exactly at the horizon ($\star F_{t\phi}=0$), both the ratio $q=F_{t\phi}/\star F_{t\phi}$ and thus $\mathcal{Z}_0$ (Equation~\eqref{eq:Z0}) are indeterminate. Finding the value of $\mathcal{Z}$ at the horizon in this case requires expanding the Penrose-Walker constant to subleading order in $\Delta$, and we find that $\mathcal{Z}$ in this case still depends on the field structure and does not converge to the universal value (Equation~\eqref{eq:Z0}). See Appendix~\ref{app:npderivationstataxi}.} and the degenerate field is stationary and axisymmetric (Equation~\eqref{eq:Ftphi0}), the ratio $\mathcal{Z}_0$ takes on a simple form that is \emph{independent of all parameters defining the electromagnetic field} and depends only on the black hole spin, observer inclination, and screen position $(\alpha,\beta)$ inside the critical curve. It is:
\begin{equation}
\label{eq:Z0}
   \mathcal{Z}_0 = \frac{\beta_{\rm s}}{\nu_{\rm s}}.
\end{equation}
In Section~\ref{sec:discussion}, we provide a physical interpretation of this universality.

Using Equation~\eqref{eq:zdef}, the horizon value of $z=\kappa_1/\kappa_2$ for stationary, axisymmetric degenerate fields is
\begin{equation}
\label{eq:univvalue}
z_0 =\frac{ (a\cos\theta_{\rm s})\nu_{\rm s}+r_+\beta_{\rm s}}{r_+\nu_{\rm s}-(a\cos\theta_{\rm s})\beta_{\rm s}}.
\end{equation}
Substituting Equation~\eqref{eq:univvalue} into Equation~\eqref{eq:chiexplicit} and using the trigonometric identity for the difference of arctangents, we find the unique form of the observed EVPA for synchrotron sources asymptotically approaching the event horizon in a degenerate, stationary, axisymmetric electromagnetic field, $\chi_+$:
\begin{align}
    \label{eq:univevpa}
    \chi_+ = \arctan\left(\frac{a\cos\theta_{\rm s}}{r_+} \right) + \arctan\left(\frac{\beta_{\rm s}}{\nu_{\rm s}}\right) - \arctan\left(\frac{\beta_{\rm o}}{\nu_{\rm o}}\right),
\end{align}
where here we use explicit observer subscripts ($\nu_{\rm o}, \beta_{\rm o}$) to distinguish the image plane coordinates $\nu$ and $\beta$ from their horizon counterparts ($\nu_{\rm s}, \beta_{\rm s}$).

In Figure~\ref{fig:explainer} we plot the unique form of the EVPA in color and as integrated streamlines over the entire multiply-lensed image of the black hole inside the critical curve. We also plot the equatorial lensing bands and the sign of the outgoing $\theta$-momentum for geodesics departing the horizon for an $a_*=0.9375$ black hole viewed at an inclination of $\theta_{\rm o}=20\deg$. In Figure~\ref{fig:examplehorizons}, we plot the unique form of the EVPA (Equation~\eqref{eq:univevpa}) across several black hole spins and inclinations inside the critical curve. These images represent the unique polarized images of a black hole horizon in synchrotron radiation when the magnetosphere is infinitely conductive, time-stationary, axisymmetric, and has nonzero poloidal current.

\subsection{Comparison to Numerical Raytracing}
In Figure~\ref{fig:examplenumana}, we compare the result of our expression for the asymptotic horizon EVPA from synchrotron emission in time-stationary, axisymmetric, degenerate fields (Equation~\eqref{eq:univevpa}) with results obtained from the numerical GR radiative transfer code \texttt{ipole} \citep{Moscibrodzka2018}. In \texttt{ipole}, we consider a thin synchrotron-emitting shell of width $\Delta r= 10^{-3}r_{\rm g}$ in an axisymmetric monopole magnetosphere. We show colormaps of the EVPA of the resulting \texttt{ipole} image for an $a_*=0.9$ black hole viewed at $\theta_{\rm o}=60\deg$ inclination. In the right panel of Figure~\ref{fig:examplenumana} we plot the EVPA along contours of constant source colatitude $\theta_{\rm s}$ in the direct ($n=0$) image as a function of the normalized distance around the contour.\footnote{In Figure~\ref{fig:examplenumana} and Figure~\ref{fig:examplegrmhd} we plot EVPA as a function of the source plane distance $s$ around the contour instead of the polar angle $\varphi$ because not all contours of constant source polar angle $\theta_{\rm s}$ enclose the $\alpha=\beta=0$ origin.} The analytic formula matches the EVPA values extracted from the raytraced image along all sampled contours.

\begin{figure*}
\centering
\includegraphics[width=\textwidth]{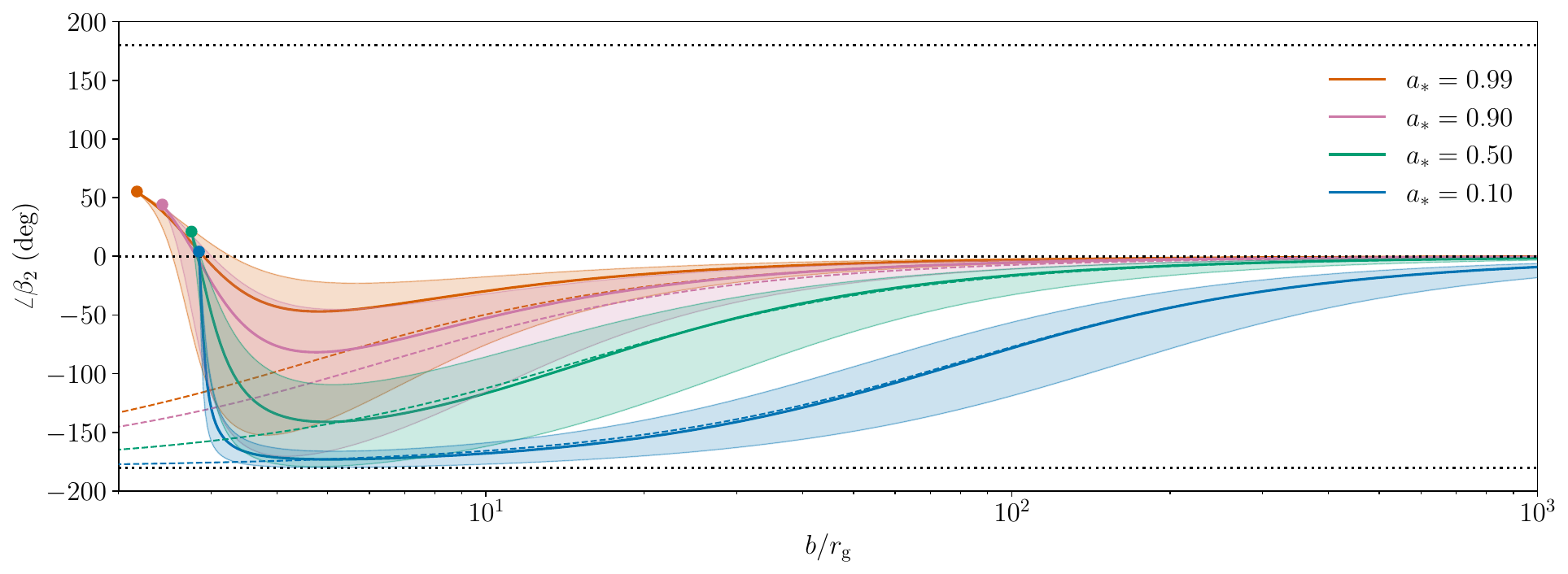}
\caption{The phase $\angle\beta_2$ as a function of observed impact parameter $b$ in an equatorial monopole model viewed face-on from the south pole ($\theta_{\rm o}=\pi$).
For several values of the black hole spin $a_*\in\{0.1,0.5,0.9,0.99\}$, we
plot $\angle\beta_2$ for models over a range of fieldline angular velocities parameterized by $w=\OmegaF/\OmegaH$. The shaded regions show the full range of $\angle\beta_2$ over $1/4\leq w \leq3/4$, while the solid lines show the models with $w=1/2$. Dashed lines indicate the observed $\angle\beta_2$ from a flat-space monopole with $w=1/2$. Solid dots at the left end of each curve show the horizon value of $\angle\beta_2$ (Equation~\eqref{eq:beta2faceon}) for face-on equatorial emission.
}
\label{fig:b2curves}
\end{figure*}
\subsection{Equatorial Emission}
We next consider the restricted case where synchrotron emitters are confined to the equatorial plane, $\theta_{\rm s}=\pi/2$.
In this case, Equation~\eqref{eq:univvalue} becomes
\begin{equation}
z_0=\frac{\pm\,\sqrt{\eta}}{\lambda-a},
\end{equation}
so the observed EVPA is
\begin{equation}
\label{eq:chiuniqueeq}
 \chi_+ = \arctan\left(\frac{\pm\sqrt{\eta}}{\lambda-a}\right) - \arctan\left(\frac{\beta_{\rm o}}{\nu_{\rm o}}\right) \;\;\; \text{(equatorial)}.
\end{equation}
This result agrees with the earlier derivation in Equations 11--13 of \citet{Hou2024}, who restricted their analysis to equatorial emission.
Note that, for emission along the black hole's equator, the sign $\pm_{\theta_{\rm s}}$ takes on a single value for each lensed image indexed by the number of equatorial crossings $n$. In particular, for $\theta_{\rm o}<\pi/2$, $\pm_{\theta_{\rm s}}=-1$ for the direct and even order images and $\pm_{\theta_{\rm s}}=+1$ for $n=1$ and all odd order images; the signs are flipped for $\theta_{\rm o}>\pi/2$.
The set of points $(\alpha,\beta)$ corresponding to direct ($n=0$) emission from the equatorial plane as $r\rightarrow r_+$ is the boundary of the black hole \emph{inner shadow} \citep{Chael2021}.

\subsection{Face-on Equatorial Emission}
Simplifying further, when an equatorial source is viewed face-on ($\theta_{\rm o} = 0$ or $\theta_{\rm o}=\pi$):
\begin{align}
\label{eq:chifaceon}
 \chi_+ &= \arctan\left(\frac{\pm\, a}{\sqrt{b^2-a^2}}\right) + \varphi,
\end{align}
where $b=\sqrt{\alpha^2+\beta^2}$ is the impact parameter on the observer screen and $\varphi=\arctan(-\alpha/\beta)$ is the observed polar angle.
Equation~\eqref{eq:chifaceon} can be simplified further by taking the Beloborodov approximation \citep{Beloborodov}, where the face-on direct image of an equatorial ring of constant $r$ appears at an impact parameter
\begin{equation}
b_{\rm bel}\approx\sqrt{r^2+2Mr+a^2}.
\end{equation}
As a result, for a face-on observer, the observed radius of the inner shadow is approximately $b_{+,\mathrm{bel}}\approx2M\sqrt{r_+/M}$, which agrees with Equation 33 of \citet{Chael2021}.
For emission at the horizon viewed face-on, the direct image value of the summary statistic $\angle\beta_2$ (Equation~\eqref{eq:beta2}) is
\begin{align}
\label{eq:beta2faceon}
\angle\beta_{2,+} &\approx 2\arctan\left(\frac{\pm\, a}{\sqrt{4Mr_+-a^2}}\right),
\end{align}
which agrees with Equation 1 of \citet{Hou2024}.

In Figure~\ref{fig:b2curves}, we illustrate the characteristic trend of $\angle\beta_2$ approaching the horizon in analytic models of a monopolar equatorial field line. We set the viewing angle to be face-on ($\theta_{\rm o}=\pi$, so the field lines rotate clockwise to match M87*; \citealt{PaperV}). We vary the field line angular velocity by sampling $w=\OmegaF/\OmegaH$ in the range $1/4\leq w \leq3/4$ and  determine the conserved poloidal current $I$ on the equatorial field line via the Znajek condition (Equation~\eqref{eq:znajek1}).
In flat space (dashed lines in Figure~\ref{fig:b2curves}), an equatorial field line with $\OmegaF\neq0$ would produce an $\angle\beta_2$ that strictly increases from $\angle\beta_2=-\pi$ (when the field lines are radial and the polarization is toroidal, e.g. at the surface of the compact object) to $\angle\beta_2=0$ in the limit $r\rightarrow\infty$ as the field lines become fully wound up and the resulting polarization becomes radial. In the Kerr monopole model, by contrast, as we approach the impact parameter of the inner shadow $b\rightarrow b_+$ from large radii, $\angle\beta_2$ reaches a minimum before turning back toward zero as the polarization again becomes more radial approaching the inner shadow. For a given black hole spin $a_*$, models spanning a range of $\OmegaF$ in our simple monopole model produce different trends of $\angle\beta_2$ and reach different minimum values at different projected radii, but they all converge to the universal horizon value of $\angle\beta_2$ (Equation~\eqref{eq:beta2faceon}) for their given black hole spin at the inner shadow.

\begin{figure*}[ht]
\centering
\includegraphics[width=\textwidth]{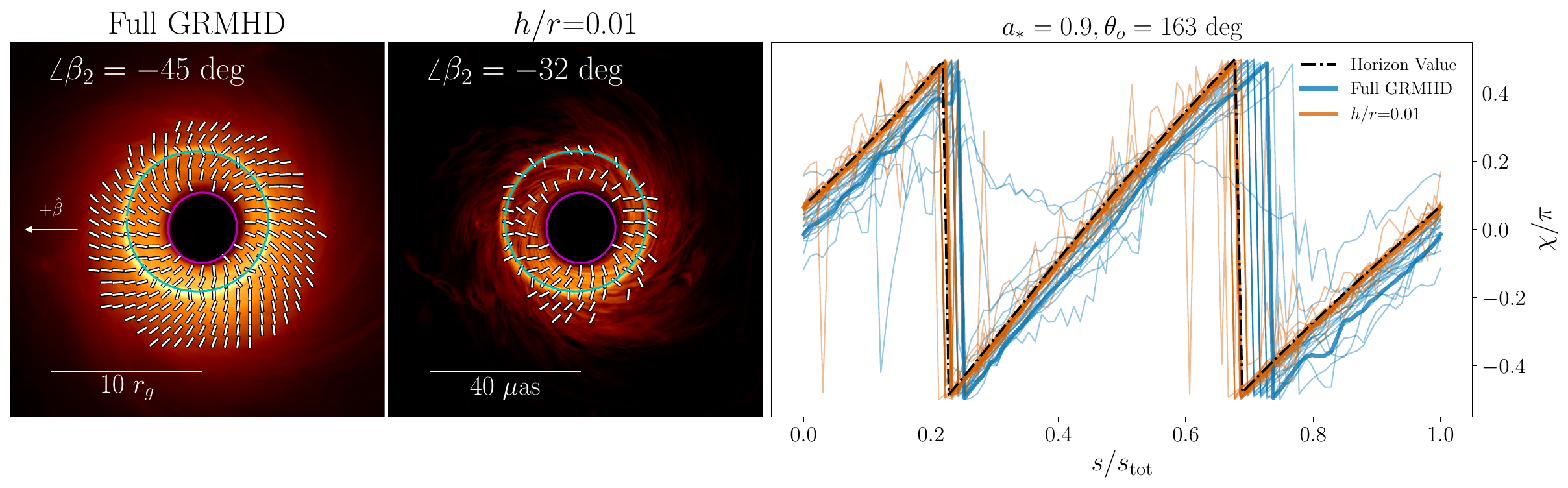}
\caption{The EVPA $\chi$ around the inner shadow taken from a MAD GRMHD simulation with spin $a_*=0.9$ and viewing angle $\theta_{\rm o}=163\deg$ and raytraced at 230 GHz with \texttt{ipole}.
\emph{Left:} The averaged GRMHD image with mass-to-distance ratio scaled to that of M87*. The inner shadow is indicated by a magenta contour while the critical curve is shown in cyan. Tick marks indicate the local averaged EVPA across the image.
\emph{Center:} The average image from raytraced snapshots where emission is restricted to lie very close to the equatorial plane, in a wedge with aspect ratio $h/r=0.01$.
\emph{Right:} The EVPAs extracted from the GRMHD snapshots (thin blue lines) and time average (thick blue lines), the corresponding EVPAs from the simulation with restricted equatorial emission (orange). The EVPAs are plotted around the inner shadow contour as a function of normalized arc length $s$ measured counterclockwise from the $+\hat{\beta}$ axis, which points to the left to match the on-sky orientation of M87*. The predicted horizon value for equatorial emission (Equation~\eqref{eq:chiuniqueeq}) is indicated by the black dashed line.}
\label{fig:examplegrmhd}
\end{figure*}

\begin{figure*}
\centering
\includegraphics[width=\textwidth]{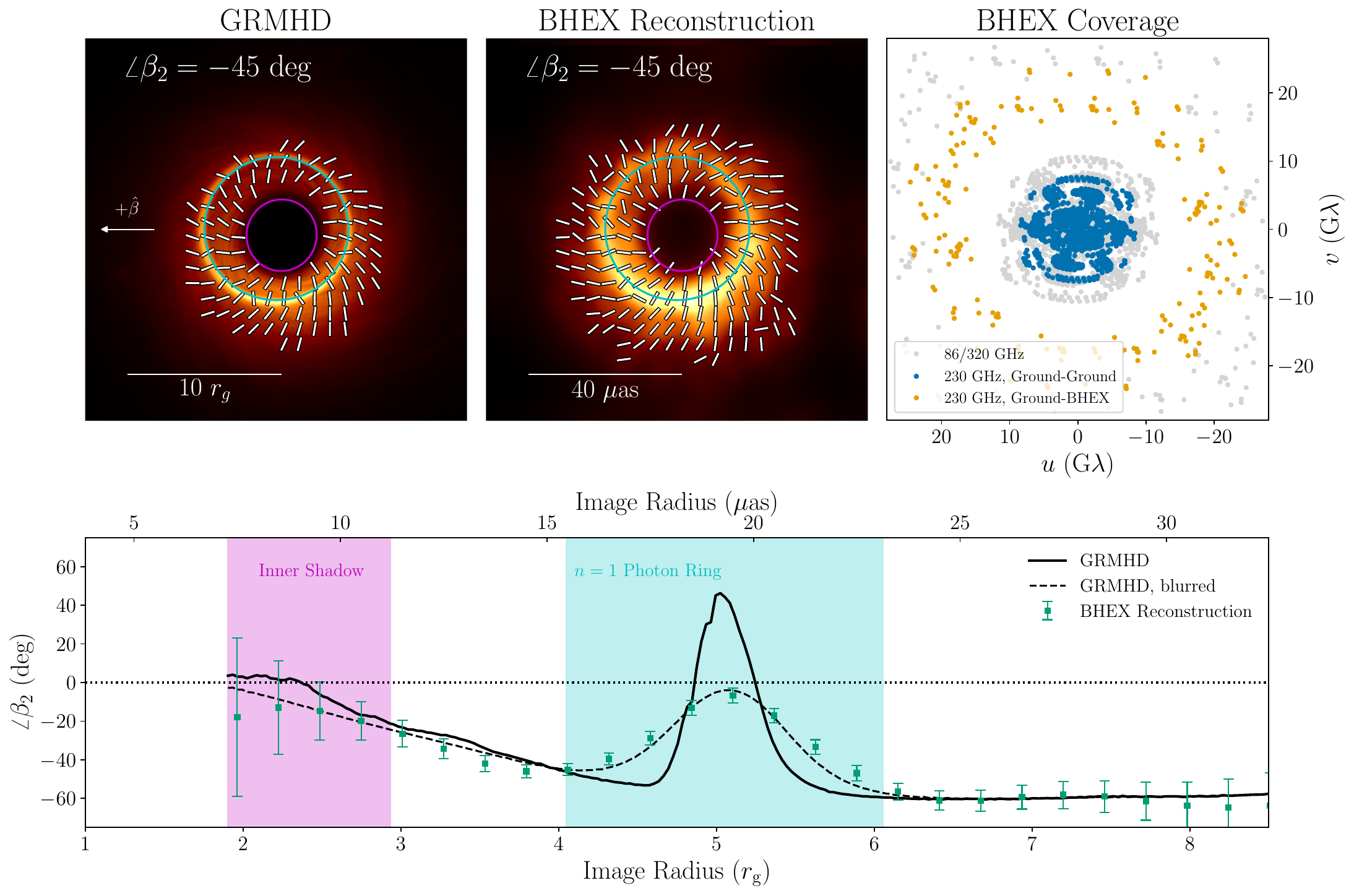}
\caption{
Potential detectability of the trend toward the universal EVPA at the horizon in M87*.
\emph{Top Left:} A time-averaged 230 GHz image of a spin $a_*=0.9$ MAD GRMHD simulation of M87* viewed at an inclination $\theta_{\rm o}=163\deg$ and rotated so the spin axis points left.
\emph{Top Center:} An idealized image reconstruction of the GRMHD image using simulated data sampled on EHT+BHEX baselines.
\emph{Top Right:} The $(u,v)$ coverage of the EHT+BHEX array used in the image reconstruction, with 230 GHz ground-ground baselines shown in blue and ground-space baselines shown in yellow. Array baselines at 86 and 320 GHz are indicated in grey.
\emph{Bottom:} Values of $\angle\beta_2$ as a function of image impact parameter extracted from the top row GRMHD image (solid black line), values extracted from the same image blurred with a 3 $\mu$as FWHM Gaussian kernel representing half of the BHEX nominal beam at 230 GHz (dashed black line), and values extracted from the reconstructed image (green squares). The $2\sigma$ error bars on the reconstruction data points were derived assuming a uniform RMS noise level of
$10^8$ K, or approximately 1\% of the peak total intensity.
}
\label{fig:examplebhex}
\end{figure*}

\section{Potential Detectability with Millimeter VLBI}
\label{sec:detection}

Here we show that in 230 GHz images from MAD simulations of the M87* black hole, there is good agreement with the universal horizon prediction in the averaged emission approaching the boundary of the inner shadow when the emission region is sufficiently compact.
While the precise horizon value at the inner shadow is subject to contamination by foreground emission and Faraday rotation,
the \emph{evolution} or \emph{swing-up} of the EVPA or $\angle\beta_2$ approaching the universal value at the inner shadow is in principle detectable with millimeter VLBI observations including a telescope in medium Earth orbit \citep{BHEX}.
Detecting the approach of the near-horizon polarization pattern toward its universal value would provide evidence that field lines thread the event horizon, a necessary condition for powering relativistic jets from spin-energy extraction.

\subsection{Universal Horizon Polarization in GRMHD}
In Figure~\ref{fig:examplegrmhd}, we plot the horizon EVPAs extracted around the $n=0$ image of the horizon in the equatorial plane --- that is, around the inner shadow --- for a series of 230 GHz images generated from a magnetically arrested (MAD) GRMHD simulation with black hole spin $a_*=0.9$. The simulation was run with the \texttt{AthenaK} code \citep{Stone2026} and raytraced with \texttt{ipole} at an inclination angle of $\theta_{\rm o}=163\deg$ with black hole mass $M$, distance $D$, and compact flux density $F_{230}$ scaled to appropriate values for M87* \citep{PaperVI}. We note that in producing the GRMHD images presented in this paper we have manually turned off the effects of Faraday rotation at 230 GHz; however, for our default choice of electron temperature parameters $R_{\rm high}=40$ and $R_{\rm low}=1$ in the \citet{Moscibrodzka2016} parameterization, we find that Faraday rotation makes a negligible contribution to the inner shadow EVPA.

The EVPA around the inner shadow in the individual snapshots as well as the regular simulation time average (blue) does not agree with the predicted universal value (Equation~\eqref{eq:chiuniqueeq}). However, when we restrict the simulation emission to lie within a thin equatorial region with aspect ratio $h/r=0.01$, the agreement between the GRMHD time average (orange) and the analytic result for stationary, axisymmetric emission (black, dashed) is striking. This is a particularly surprising result, as the synchrotron emission and parallel transport effects are nonlinear transformations of the underlying field $F_{\mu\nu}$. There is no a priori guarantee that the average image of a series of time-variable GRMHD snapshots should agree with Equation~\eqref{eq:chiuniqueeq}, which was derived assuming that the field is time-stationary and axisymmetric. The fact that the two averages do agree suggests that the fluctuations away from the condition $E^\phi=0$ at the horizon are sufficiently small that they do not bias the average of the nonlinear map from the horizon fields to the observed polarization away from the universal value. More work is required to understand in exactly what situations we may obtain the value predicted by Equation~\eqref{eq:chiuniqueeq} after time averaging a series of images from an intrinsically variable and turbulent GRMHD source.

\subsection{Observability with BHEX}
Exactly detecting the universal horizon polarization described here in an astrophysical black hole image is unlikely. Even in a perfectly axisymmetric and stationary magnetosphere, foreground emission will dominate the image over the interior of most of the critical curve; time variability and deviations from axisymmetry will also introduce deviations from the prediction of Equation~\eqref{eq:univevpa}.
However, the characteristic swing in $\angle\beta_2$ toward the inner shadow (as the EVPAs become more radial) seen in Figure~\ref{fig:b2curves} may be observable in real sources where the near-horizon emission is concentrated in the equatorial plane, as it is in MAD GRMHD simulations (\citealt{PaperV}, \citetalias{BHPII}).
Indeed, in time-averaged GRMHD images we find that while the average $\angle\beta_2$ trend with impact parameter $b$ does not reach the predicted inner shadow value, we nonetheless still observe a characteristic upward swing toward more radial EVPA structure ($\angle\beta_2 \rightarrow 0$) at smaller impact parameters inside the photon ring (black curve in Figure~\ref{fig:examplebhex}).

In Figure~\ref{fig:examplebhex} we perform a proof-of-concept reconstruction of a time-averaged GRMHD simulation of M87* using simulated data from a 230 GHz VLBI array including an observatory in medium Earth orbit.  We model our source after M87* by adopting a mass $M=6.5\times10^9\,M_\odot$ and distance $D=16.8$ Mpc \citep{PaperVI}, thus scaling its apparent angular size to $r_{\rm g}/D=3.82\,\mu$as. The simulation is observed at an inclination angle $\theta_{\rm o}=163\deg$ and the on-sky position angle is oriented so the projected spin axis points East; we scale the total flux density at 230 GHz to $F_{230}=0.5$ Jy \citep{PaperIV, EHT_2021}. Our array is modeled after the proposed Black Hole Explorer (BHEX) mission \citep{BHEX}; its Earth-space baselines enable a factor of $3\times$ higher resolution than the current ground-based EHT at 230 GHz. We generate synthetic data from the time-averaged GRMHD simulation observed with this array using the \texttt{ngehtsim} code \citep{Pesce2024}; we then reconstruct an image from the synthetic data using the \texttt{eht-imaging} code \citep{Chael2016,Chael2018}.

After reconstructing an image, we extract values of $\beta_2$ on circular annuli from the ground truth image and the reconstruction using Equation~\eqref{eq:beta2}; to counteract the spin-induced shift of the ring centroid from the $\alpha=\beta=0$ origin, we manually shift all images by a displacement of $-2Ma_*\sin\theta_{\rm o}/D$ \citep{Chael2021}.
Our synthetic data and imaging procedure are highly idealized; they assume no systematic calibration errors either in the individual telescope gains or complex polarization leakage terms. Nonetheless, the example in Figure~\ref{fig:examplebhex} shows that 230 GHz VLBI observations of M87* that combine high-resolution ground-space baselines with high dynamic range baselines in a dense ground array may be able to detect the characteristic swing in the near-horizon polarization angle
toward the inner shadow. Future work that accounts for more realistic corruptions in synthetic data as well as the effects of non-zero inclination on the predicted $\angle\beta_2$ trend will be necessary to make a robust prediction for the observability of this signal with BHEX and its potential to constrain the black hole spin.

\section{Discussion}
\label{sec:discussion}
\subsection{Physical Interpretation}
In Appendix~\ref{app:zamoderivation}, we re-derive Equation~\eqref{eq:Z0general} to determine the asymptotic horizon polarization for synchrotron emitters in a degenerate electromagnetic field working entirely in the Boyer-Lindquist normal-observer frame. This derivation is more complicated than our initial derivation in the NP formalism in Appendix~\ref{app:npderivation}; however, it provides more physical insight. We show that in this picture, the emergence of a universal horizon polarization for synchrotron radiation when $E^\phi=0$ is a consequence of the source frame polarization vector $f^{(\hat a)}$ becoming aligned with the polar direction $(\hat\theta)$ to leading order. Approaching the horizon, the outgoing photon momentum diverges and becomes
predominantly radial while the emitting plasma is dragged inward. Meanwhile, the stationary, axisymmetric magnetic field in the emitter's frame becomes purely toroidal. The synchrotron polarization is then left with only the polar direction available, independent of the
details of the field.

The ZAMO frame has a clean relationship to the coordinate frame and preserves the orientation of the polar direction $\hat\theta$ everywhere; however, it has the notable disadvantage of becoming null on the horizon, which introduces multiple divergences into our derivation in Appendix~\ref{app:zamoderivation}.
In Appendix~\ref{app:driftderivation} we re-derive the result yet again by working in the plasma drift frame, where all divergences are traceable to the blow-up of the local frame photon energy $E'$. We find in this derivation that the stationary, axisymmetric magnetic field becomes entirely aligned with the projection of the global $\hat\phi$ direction into the drift frame; the orthogonal drift frame polarization vector $f^{(\hat a)}$ is thus forced into the projection of the global $\hat\theta$ direction. In both of these calculations, we show that a nonzero toroidal electric field ($q\neq0$) induces a leading-order polar component in the magnetic field and spoils the universality (Equation~\eqref{eq:Z0}); universality is also lost in the edge case when the toroidal magnetic field exactly vanishes ($q$ is undefined).

We thus conclude that the unique asymptotic EVPAs from synchrotron emitters approaching the horizon in a stationary, axisymmetric, degenerate magnetosphere in some sense produce images of the black hole's lines of longitude. At infinity, we observe the local polar direction on the black hole horizon, twisted and distorted on its way to the observer by the effects of strong gravitational lensing and spin-dependent frame-dragging.

\subsection{Relation to \citet{Lupsasca2020}}
Intriguingly, \citet{Lupsasca2020} found a result that at first glance appears similar to our result in Equation~\eqref{eq:univvalue} for highly spinning black holes $a_*\approx1$. They considered polarization vectors $f^\mu$ that are invariant under the expanded class of symmetries that emerge in the throat region of near-extremal black holes and found these polarization vectors source a Penrose-Walker constant with ratio $z=\kappa_1/\kappa_2=\beta_{\rm s}/\nu_{\rm s}$. While the result of \citet{Lupsasca2020} looks superficially similar to the result derived here for $\mathcal{Z}=\beta_{\rm s}/\nu_{\rm s}$ across all black hole spins, these results are in fact related non-trivially by Equation~\eqref{eq:zdef} and will yield different observed EVPAs. The structural resemblance in our results may reflect a deeper connection between the symmetries of the near-extremal black hole polarization exploited by \citet{Lupsasca2020} and the general conditions on synchrotron radiation and degenerate fields that drive our result, but we leave further exploration of this connection to future work.

\section{Conclusion}
\label{sec:conclusions}
In this paper, we have shown that for degenerate electromagnetic fields around a Kerr black hole with nonzero toroidal Boyer-Lindquist magnetic field but vanishing toroidal electric field --- a class which includes all axisymmetric, time-stationary, degenerate field configurations with nonzero poloidal current --- the observed EVPA from a synchrotron emitter approaching the event horizon takes on a unique, field-independent value $\chi_+$. Neglecting the effects of Faraday rotation and foreground emission, this value depends only on the spin of the black hole $a$, the observer inclination $\theta_{\rm o}$, and the Bardeen coordinates $(\alpha,\beta)$ within the black hole critical curve for the null geodesic connecting the source at the horizon to the observer at infinity. Furthermore, $\chi_+$ can be written in an exceptionally simple form (Equation~\eqref{eq:univevpa}) in terms of the shifted Bardeen coordinates $(\nu,\beta)$ and their analogues on the horizon $(\nu_{\rm s}, \beta_{\rm s})$. In the specific case of direct $n=0$ emission from the equatorial plane (Equation~\eqref{eq:chiuniqueeq}), our result reproduces the earlier result of \citet{Hou2024}.

In Figure~\ref{fig:examplegrmhd}, we find that in 230 GHz images from full 3D GRMHD simulations of M87* with non-axisymmetric and time-variable emission,
the average value of the EVPA around the inner shadow approaches the universal value for equatorial emission predicted in Equation~\eqref{eq:chiuniqueeq}
as long as Faraday
rotation is minimal and the emitting region is taken to be sufficiently narrow in the equatorial plane.
While it is unlikely that the unique horizon polarization signature we derived can be detected in any astrophysical black hole image, we show in Figure~\ref{fig:examplebhex} that the swing of the EVPAs toward the radial direction for equatorial emission approaching the inner shadow may be detectable in M87* by the space-based VLBI experiment BHEX \citep{BHEX}.
Detecting this swing in M87* would be a strong indication of both the winding up of magnetic fields and strong parallel transport of polarization close to the horizon of a black hole; it would indicate that the observed emission comes from magnetic field lines that thread the event horizon, which is a requirement for \citet{BlandfordZnajek1977} spin-energy extraction. Connecting the polarization evolution observed at small apparent radii to its expected swing back toward the radial direction at larger radii in the downstream jet (at the black hole light cylinder; \citealt{Gelles2025,Gelles2026}) could further provide evidence that magnetic field lines in the jet extend to the black hole horizon.

More work is needed to develop robust multi-scale fittable physical models of the emission region to measure the key parameters that set the average Poynting flux -- the magnetic field strength $|B|$, the ratio $B^\phi/B^r$, and rotation rate $\OmegaF$ -- using polarimetric VLBI data at different radii. Measuring $\OmegaF$ is particularly important, as in BZ models the factor $\OmegaF-\OmegaH$ determines the sign and speed of the pitch angle evolution (Figure~\ref{fig:b2curves}). In particular, if the M87* accretion flow is retrograde, as suggested in some interpretations of EHT data \citep{PaperVIII,Qiu2023}, constraining the sign of $\OmegaF$ to be counteraligned with the large-scale accretion flow would provide strong evidence that the field line rotation is set by the black hole spin \citep[see e.g.][]{Ricarte2022,Conroy2023,Conroy2026}.

While exceptional care would need to be taken to properly calibrate the effects of Faraday rotation and off-equatorial emission geometry, a measurement of the rate of the linear polarization swing toward more radial values at the horizon with high-resolution space VLBI experiments may be used to constrain the energy flux near the black hole and potentially even the black hole spin.
Observing this swing-up in the EVPA toward the universal average value described in this paper would probe both the wind-up of magnetic field lines from a spinning black hole and the spin-dependent parallel transport of light in the strongly curved spacetime near the event horizon. Combined with multi-scale data from VLBI images at different radio frequencies, constraining the evolution of polarization toward its universal value at the horizon may link the extragalactic jet in M87* to horizon-threading magnetic field lines that extract electromagnetic energy from the spinning black hole.

\software{\texttt{kgeo} \citep{Chael2021, kgeo},
\texttt{AthenaK} \citep{Stone2026},
\texttt{ipole} \citep{Noble2007,Moscibrodzka2018},
\texttt{eht-imaging} \citep{Chael2016,Chael2018,eht-imaging},
\texttt{ngehtsim} \citep{Pesce2024},
NumPy \citep{NumPy2020},
SciPy \citep{SciPy2020},
Matplotlib \citep{Matplotlib}}

\begin{acknowledgements}
We thank Michael Johnson, Ramesh Narayan, and Gautam Satishchandran for insightful conversations. Daniel Palumbo and Dom Pesce provided the BHEX $(u,v)$ coverage used in Figure~\ref{fig:examplebhex}.
AL was supported in part by NSF grant AST-2307888, the NSF CAREER award PHY-2340457, and the Simons Foundation grant SFI-MPS-BH-00012593-09.
AC was supported by the Villum Fonden
grant No. 82533, and by the Princeton Gravity Initiative.
GNW was supported by the Taplin Fellowship and by the Princeton Gravity Initiative. ZG was supported by an NSF Graduate Research Fellowship.
EQ was supported in part by a Simons Investigator award from the Simons Foundation.
AL and EQ thank the Aspen Center for Physics, which is supported by NSF grant PHY-1607611. The authors used Claude Opus 4.8 (Anthropic) and GPT-5.5 Pro (OpenAI) to check grammar and formatting, to assist with coding inquiries, and to validate equations symbolically and numerically. All output was reviewed and verified by the authors, who take full responsibility for the content of this manuscript.

\end{acknowledgements}

\newpage
\def\subsectionautorefname{section}
\appendix
\section{Review of essential properties of Kerr}
\label{app:KerrSymmetry}

In this Appendix, we review several essential properties of the Kerr spacetime required in the derivation of our key result for the limiting horizon EVPA value for synchrotron radiation in stationary, axisymmetric, degenerate magnetospheres in Equation~\eqref{eq:univevpa}.

\subsection{Conformal Killing-Yano Tensors and Conservation of the Penrose-Walker Constant}
In this section, we review how the conservation of the Penrose-Walker constant follows directly from the properties of the two conformal Killing-Yano tensors in the Kerr metric.
A Killing-Yano (KY) tensor is a rank-2 antisymmetric tensor that obeys the Killing-Yano equation:
\begin{align}
	\label{eq:KY}
	\nabla_{(\lambda}J_{\mu)\nu}=0.
\end{align}
A KY tensor can always be ``squared'' to obtain a Killing tensor
\begin{align}
\label{eq:KYsq}
	K_{\mu\nu}=-{J_\mu}^\lambda J_{\lambda\nu},
\end{align}
which obeys the Killing equation, $\nabla_{(\lambda}K_{\mu\nu)}=0.$
\emph{Conformal} Killing-Yano tensors (CKYs; \citealt{Tachibana}) extend this definition and obey the equation
\begin{align}
	\label{eq:CKY}
	\nabla_{(\lambda}Q_{\mu)\nu}=V_\nu \,g_{\lambda\mu}-V_{(\lambda}\, g_{\mu)\nu}.
\end{align}
In Equation~\eqref{eq:CKY}, $V^\mu=\frac{1}{3}\nabla_\nu Q^{\nu\mu}$ is a divergence-free vector associated with the CKY tensor $Q_{\mu\nu}$.

One can show that if $g_{\mu\nu}$ solves the vacuum Einstein equations and $Q_{\mu\nu}$ is a CKY tensor, then its associated vector $V^\mu$ is a Killing vector field of the metric $g_{\mu\nu}$ \citep[][Theorem 4]{Jezierski}. One can further show that in a four-dimensional spacetime, $Q_{\mu\nu}$ is a CKY tensor if and only if its Hodge dual $\star Q_{\mu\nu}$ is also a CKY tensor \citep[][Theorem 5]{Jezierski}. The Kerr metric admits a single such pair.
The exact Killing-Yano tensor in Kerr is
\begin{equation}
\label{eq:KYBL}
J=r\sin{\theta}\ed\theta\wedge\br{\pa{r^2+a^2}\ed\phi-a\ed t}-a\cos{\theta}\ed r\wedge\pa{a\sin^2{\theta}\ed\phi-\ed t}.
\end{equation}
The Hodge dual of the exact Kerr KY tensor $J_{\mu\nu}$ is a CKY tensor:
\begin{equation}
\label{eq:CKYBL}
\Jd=a\cos\theta\sin\theta\,\ed\theta\wedge\br{\pa{r^2+a^2}\ed\phi-a\ed t}+r\ed r\wedge\pa{a\sin^2{\theta}\ed\phi-\ed t}.
\end{equation}
The associated vector to $\Jd$ is the time translation Killing vector $V=\pd_t$.

The conservation of the Penrose-Walker constant (Equation~\eqref{eq:KYPW}) follows directly from the conformal Killing-Yano equation (Equation~\eqref{eq:CKY}) and the orthogonality and parallel transport of $f^\mu$ (Equation~\eqref{eq:PolarizationConditions}). Starting with the real part of the Penrose-Walker constant, $\kappa_1=\Jd_{\alpha\beta}p^\alpha f^\beta$ and taking its derivative along a photon trajectory with momentum $p^\mu$:
\begin{align}
p^\mu\nabla_\mu\,\kappa_1 &= \Jd_{\alpha\beta} \left(f^\beta p^\mu \nabla_\mu p^\alpha +  p^\alpha p^\mu \nabla_\mu f^\beta\right) + p^\alpha f^\beta  p^\mu \nabla_\mu \Jd_{\alpha \beta}, \nonumber \\
&= f^\mu p^\alpha p^\beta \nabla_{(\alpha} \Jd_{\beta)\mu}, \nonumber \\
&= \frac{1}{2} f^\mu p^\alpha p^\beta \left(2V_\mu g_{\alpha\beta}-V_\alpha g_{\mu\beta} - V_\beta g_{\mu\alpha}\right), \nonumber \\
&= (V \cdot f)(p\cdot p) - (V\cdot p)(p\cdot f), \nonumber \\
&= 0.
\end{align}
The first and second terms in the first line vanish by the parallel transport of both $p^\alpha$ and $f^\beta$ along the geodesic. In the second line we symmetrize over $p^\alpha p^\beta$, and in the third line we substitute in the CKY equation (Equation~\eqref{eq:CKY}). The first term in the fourth line vanishes because $p^\mu$ is null and the second term vanishes because $p\cdot f=0$.
The same argument applies for the conservation of the imaginary part of the Penrose-Walker constant $\kappa_2=J_{\mu\nu}p^\mu f^\nu$, except that the third line is identically zero by the KY equation (Equation~\eqref{eq:KY}).

\subsection{Newman-Penrose Null Basis}
\label{app:NPdefs}
Here we review how vectors and tensors are expressed in a principal null basis in the Newman-Penrose formalism.
\subsubsection{Basis Vectors}
We work with the Hartle-Hawking null tetrad $e^{\mu}_a=\left(\ell^\mu, n^\mu, m^\mu, \overline{m}^\mu\right)$, where $\ell^\mu$ is the principal outgoing null vector, $n^\mu$ is the principal ingoing null vector, and $\overline{m}^\mu$ and $m^\mu$ are complex null vectors that together encode the angular directions.
The basis vectors have Boyer-Lindquist  components:
\begin{subequations}
\label{eq:nulltetrad}
\begin{align}
\ell^\mu &= \frac{1}{2(r^2+a^2)}\left(r^2+a^2,\Delta,0,a\right),\\
n^\mu &= \frac{(r^2+a^2)}{\Delta\Sigma}\left(r^2+a^2,-\Delta,0,a\right), \\
m^\mu &= \frac{1}{\sqrt{2}(r+ia\cos\theta)}\left(ia\sin\theta,0,1,i\csc\theta\right), \\
\overline{m}^\mu &= \frac{1}{\sqrt{2}(r-ia\cos\theta)}\left(-ia\sin\theta,0,1,-i\csc\theta\right).
\end{align}
\end{subequations}
The basis vectors obey the orthonormality properties $\ell\cdot n=-1$ and $m\cdot\overline{m}=1$, while $\ell\cdot\ell=n\cdot n = m \cdot m = \overline{m}\cdot\overline{m}= \ell\cdot m = n\cdot m = \ell \cdot \overline{m} = n\cdot\overline{m}=0$.

We note that we choose to use the Hartle-Hawking form of the null tetrad (Equation~\eqref{eq:nulltetrad}) rather than the standard Kinnersley form \citep{Kinnersley}. These forms are related by a rescaling in the ingoing-outgoing sector: $\ell^{\mu}=\frac{\Delta}{2(a^2+r^2)}\ell^{\mu}_{\mathrm{K}}$, $n^{\mu}=\frac{2(r^2+a^2)}{\Delta}n^{\mu}_{\mathrm{K}}$ \citep[e.g.][]{Poisson}.
We choose the Hartle-Hawking tetrad because it is regular on the future horizon, which we can see by expressing $\ell$ and $n$ in ingoing coordinates $\mathrm{d}v=\mathrm{d}t + (r^2+a^2)\mathrm{d}r/\Delta$ and $\mathrm{d}\tilde{\phi}=\mathrm{d}\phi + a\,\mathrm{d}r/\Delta$:\footnote{Note that in the GRMHD literature, the Kerr-Schild ingoing coordinates are typically defined with $\mathrm{d}v=\mathrm{d}t+2Mr\,\mathrm{d}r/\Delta$ \citep[e.g.][]{McKinney2004}. The outgoing null vector $\ell$ is still manifestly regular on the horizon in Kerr-Schild coordinates, but it is messier than using the normalization chosen here.}
\begin{subequations}
\label{eq:nlinfalling}
\begin{align}
\ell &= \partial_v + \frac{\Delta}{2(r^2+a^2)}\partial_r + \frac{a}{r^2+a^2}\partial_{\tilde{\phi}}, \\
n &= -\frac{(r^2+a^2)}{\Sigma}\partial_r.
\end{align}
\end{subequations}
At the future horizon, $\mathcal{H}_+$, the outgoing $\ell = \partial_v + \OmegaH\partial_{\tilde{\phi}}$ reduces to the horizon generator, while the ingoing $n = -(2Mr_+/\Sigp) \partial_r$ points to the black hole interior. In the standard Kinnersley tetrad, $\ell_{\rm K}$ and $n_{\rm K}$ are instead regular on the \emph{past} horizon $\mathcal{H}_-$, where $n_{\rm K}$ becomes the horizon generator and $\ell_{\rm K}$ points to the black hole exterior.

\subsubsection{Maxwell-Newman-Penrose Scalars}
In the Newman-Penrose formalism, the six degrees of freedom in the electromagnetic field $F$ are typically expressed in the three complex Maxwell-Newman-Penrose scalars \citep{Teukolsky}:
\begin{subequations}
\label{eq:npscalars}
\begin{align}
\Phi_0 &= F_{\ell m} = F_{\mu\nu}\ell^\mu m^\nu, \\
\Phi_1 &= \frac{1}{2}\left(F_{\ell n} + F_{\overline{m} m}\right) = \frac{1}{2}F_{\mu\nu}\left(\ell^\mu n^\nu + \overline{m}^\mu m^\nu\right), \\
\Phi_2 &= F_{\overline{m} n} = F_{\mu\nu}\overline{m}^\mu n^\nu,
\end{align}
\end{subequations}
so that:
\begin{equation}
F_{\mu\nu} = 2\left[\Phi_1\left(n_{[\mu}\ell_{\nu]}+m_{[\mu}\overline{m}_{\nu]}\right)+\Phi_2\ell_{[\mu}m_{\nu]}+\Phi_0\overline{m}_{[\mu}n_{\nu]}\right] + \mathrm{c.c.}\;,
\end{equation}
where ``c.c.'' indicates the complex conjugate. We will show in Appendix~\ref{app:npderivation} that the Maxwell-Newman-Penrose decomposition is particularly suited for describing the polarization of synchrotron radiation in a degenerate electromagnetic field.

\subsubsection{Kerr Killing-Yano Tensors in Null Coordinates}
The symmetries of the conformal Killing-Yano tensors $J_{\mu\nu}$ and $\Jd_{\mu\nu}$ in Kerr are made manifest when expressing them in the null basis, where they take on a block-diagonal form.
The Kerr CKY and KY tensors are
\begin{equation}
\label{eq:KYnull}
J_{\mu\nu} = 2\left(-a\cos\theta\,\ell_{[\mu}n_{\nu]}+ir\,m_{[\mu}\overline{m}_{\nu]}\right) \quad,\quad
\Jd_{\mu\nu} = 2\left(r\,\ell_{[\mu}n_{\nu]}+ia\cos\theta \, m_{[\mu}\overline{m}_{\nu]}\right).
\end{equation}
Thus the complex tensor $\Jd_{\mu\nu} + iJ_{\mu\nu}$ neatly factors into ingoing-outgoing and angular parts:
\begin{equation}
\label{eq:complexJ}
\Jd_{\mu\nu} + iJ_{\mu\nu} = 2(r-ia\cos\theta)\left(\ell_{[\mu}n_{\nu]}  - m_{[\mu}\overline{m}_{\nu]} \right).
\end{equation}
The real $\mathcal{A}$ and $\mathcal{B}$ parts of the PW constant (Equation~\eqref{eq:ABdef}) are thus
\begin{subequations}
\label{eq:ABnp}
\begin{align}
	\mathcal{A}&= 2\,\ell_{[\mu}n_{\nu]}p^\mu f^\nu \label{eq:Anp}, \\
	\mathcal{B}&= -2i\,m_{[\mu}\overline{m}_{\nu]}p^\mu f^\nu \label{eq:Bnp},
\end{align}
\end{subequations}
or, in terms of vector components in the null basis:
\begin{equation}
\label{eq:AmBnull}
\mathcal{A}-i\mathcal{B} = \left(p^n f^\ell - p^\ell f^n\right) + \left(p^m f^{\overline{m}} - p^{\overline{m}}f^m\right).
\end{equation}
Equation~\eqref{eq:AmBnull} shows that $\mathcal{A}$ physically corresponds to the projection of the plane of polarization $p\wedge f$ into the null radial plane $\ell \wedge n$, while $\mathcal{B}$ corresponds to the projection of the plane of polarization into the angular plane $m\wedge \overline{m}$.

\section{Derivation of the Horizon Value $\mathcal{Z}_0$ in the Newman-Penrose Formalism}
\label{app:npderivation}
In this Appendix we derive the equation for the unique horizon polarization $\mathcal{Z}_0$ using the Newman-Penrose formalism. We first show that the Penrose-Walker constant for synchrotron emitters approaching the horizon takes on a simple form for all degenerate electromagnetic fields. We then show that when applying the horizon boundary conditions on the Boyer-Lindquist field components $E^i$ and $B^i$, the observed polarization direction from the horizon takes on a simple form for arbitrary emitters in a magnetically dominated degenerate field. Finally, we specialize to the stationary, axisymmetric case and show that all dependence on the form of the electromagnetic field drops out of the observed horizon polarization in this case as a result of the vanishing toroidal electric field $E^\phi=0$.
\subsection{Photon Momentum}
First, we inspect the components of the outgoing ($\pm_r=+1$) photon momentum vector $p^a$ in the null basis:
\begin{subequations}
\begin{align}
\label{eq:pell}
p^\ell &= -n^\mu p_\mu =  E\frac{r^2+a^2}{\Delta\Sigma}\left(r^2+a^2-a\lambda+\sqrt{\mathcal{R}(r)}\right),\\
\label{eq:pn}
p^n &= -\ell^\mu p_\mu =  E \frac{1}{2(r^2+a^2)}\left(r^2+a^2-a\lambda-\sqrt{\mathcal{R}(r)}\right),\\
\label{eq:pm}
p^m &= \bar{m}^\mu p_\mu =  E \frac{\pm_\theta \sqrt{\Theta(\theta)}-i(\lambda\csc\theta-a\sin\theta)}{\sqrt{2}\,\Sigma}\,\left(r+ia\cos\theta\right).
\end{align}
\end{subequations}
It is apparent from Equation~\eqref{eq:pell} that the component of the photon momentum along the outgoing null geodesic, $p^\ell$, diverges at the horizon:
\begin{equation}
    p^\ell = \Delta^{-1}p^\ell_{-1} + p^\ell_0 + \Delta p^\ell_1 + \ldots \;.
\end{equation}
By substituting Equation~\eqref{eq:Rpot} into Equation~\eqref{eq:pn} we can see that the component of the photon momentum along the ingoing null geodesic vanishes at the horizon, $\lim_{\Delta\rightarrow 0} p^n = 0$, so that
\begin{equation}
\label{eq:pnvanish}
    p^n = \Delta p^n_1 + \Delta^2 p^n_2 + \ldots \;.
\end{equation}
Both $p^m$ and $p^{\overline{m}}$ are finite but nonzero on the horizon. Since $p^2=0$ for null geodesics, the components in the null basis are related by:
\begin{equation}
\label{eq:nullprelation}
p^\ell p^n = p^m p^{\overline{m}}
= |p^m|^2.
\end{equation}
This equation requires that $p^n\propto\Delta\rightarrow 0$ at the horizon in order to keep the angular part $|p^m|^2$ of the photon momentum finite as the outgoing component $p^\ell$ diverges as $1/\Delta$.
The fact that the $p^\ell$ component of an outgoing null vector diverges and the $p^n$ component vanishes is physically interpretable as the fact that no outgoing null rays from exactly on the horizon can physically reach the observer at infinity.

\subsection{Synchrotron Polarization}
Since we demand the components of any physical electromagnetic field $F$ to be finite for infalling observers, the components of $F$ in the Hartle-Hawking null basis (Equation~\eqref{eq:npscalars}) should also be regular on the horizon.
Synchrotron emission from a degenerate field has the unique property that its polarization is obtained by contracting $F$ with $p$ (Equation~\eqref{eq:polvec2}):
\begin{equation}
f^a = F^a_{\;\;b\,}p^b.
\end{equation}
This means that the Penrose-Walker constant (Equation~\eqref{eq:KYPW}) is bilinear in $p$:
\begin{equation}
(\mathcal{A}-i\mathcal{B}) = p^a\left(W_{ab}\, g^{bc} F_{cd}\right) p^d,
\end{equation}
where we define the reduced complex CKY tensor as $W_{ab}\equiv (r-ia\cos\theta)^{-1}(\Jd + iJ)_{ab}$. The matrix forms of the metric $g_{ab}$, the reduced complex CKY tensor $W_{ab}$, and the field strength tensor $F_{cd}$ are
\begin{equation}
g_{ab} = \begin{pmatrix}
0 & -1 & 0 & 0 \\
-1 & 0 & 0 & 0 \\
0 &  0 & 0 & 1 \\
0 &  0 & 1 & 0
\end{pmatrix}, \qquad
W_{ab} = \begin{pmatrix}
0 & -1 & 0 & 0 \\
1 &  0 & 0 & 0 \\
0 &  0 & 0 & 1 \\
0 &  0 &-1 & 0
\end{pmatrix}, \qquad
F_{cd} = \begin{pmatrix}
0 & 2\mathrm{Re}[\Phi_1] & \Phi_0 & \overline{\Phi}_0 \\
-2\mathrm{Re}[\Phi_1] & 0 & -\overline{\Phi}_2 & -\Phi_2 \\
-\Phi_0 & \overline{\Phi}_2 & 0 & -2i\,\mathrm{Im}[\Phi_1] \\
-\overline{\Phi}_0 & \Phi_2 & 2i\,\mathrm{Im}[\Phi_1] & 0
\end{pmatrix}.
\end{equation}
Thus, using $p^2=0$ we find that:
\begin{equation}
\label{eq:niceAB}
\mathcal{A}-i\mathcal{B} = 2\left(\overline{\Phi}_0p^{\overline{m}}p^\ell + \overline{\Phi}_2 p^m p^n + 2\overline{\Phi}_1|p^m|^2\right).
\end{equation}
This formula is exact and holds for all synchrotron radiation from degenerate fields in Kerr and does not depend on assuming the field is stationary and axisymmetric.

\subsection{Synchrotron Polarization at the Horizon}
Equation~\eqref{eq:niceAB} makes the asymptotic behavior of $\mathcal{A}-i\mathcal{B}$ for observers approaching the horizon obvious. We have already established that the only diverging part of either $p^a$ or $F_{ab}$ as $\Delta\rightarrow0$ is $p^\ell$, the component of the photon momentum along the principal outgoing null congruence. As a result, at the horizon $\mathcal{A}-i\mathcal{B}$ diverges as $\Delta^{-1}$, with a leading order coefficient:
\begin{equation}
\label{eq:ABdiv}
\lim_{\Delta\rightarrow 0} \Delta \left(\mathcal{A}-i\mathcal{B}\right) = 2 p^\ell_{-1} p^{\overline{m}} \overline{\Phi}_0.
\end{equation}
Therefore, the horizon limit of $\mathcal{Z}=-\mathcal{A}/\mathcal{B}$ is
\begin{equation}
\label{eq:Z0nullgen}
    \mathcal{Z}_0 = -\left.\frac{\mathrm{Re}\left[p^m\Phi_0\right]}{\mathrm{Im}\left[p^m\Phi_0\right]}\right|_{r=r_+}.
\end{equation}
The simple form of $\mathcal{Z}_0$ at the horizon does not rely on the field $F_{ab}$ being axisymmetric and time-stationary; these equations hold for any synchrotron emitter in a degenerate EM field, as long as we can assume that the Newman-Penrose field scalars $\Phi_0,\Phi_1,\Phi_2$ are regular on the horizon. It is apparent from the form of Equation~\eqref{eq:Z0nullgen} that the necessary and sufficient condition for a synchrotron emitter in a degenerate field to have an asymptotic horizon EVPA that is independent of the specific electromagnetic field is that the phase of $\Phi_0$ be independent of the field.

\subsection{Form of $\Phi_0$}

Next, we consider the form of the Newman-Penrose scalar $\Phi_0$ in Kerr, both generally and in the specific case of axisymmetric, time-stationary degenerate fields. After writing down a general Kerr $F_{\mu\nu}$ in terms of the BL coordinate frame fields $E^i = F^{ti} , B^i=\star F^{it}$, we find from the definition $\Phi_0 = \ell \cdot F \cdot m$ that
\begin{subequations}
\label{eq:phi0}
\begin{align}
\Phi_0 &= \frac{\sin\theta}{2\sqrt{2}}\frac{(r-ia\cos\theta)}{(r^2+a^2)}\left(X+iY\right), \\
X&\equiv-\frac{\Sigma\Delta}{\Pi}\left(aB^r + (r^2+a^2)\csc\theta\,E^\theta\right) + \Delta B^\phi, \\
Y&\equiv\hphantom{-}\frac{\Sigma\Delta}{\Pi}\left(aE^r - (r^2+a^2)\csc\theta\,B^\theta\right) - \Delta E^\phi,
\end{align}
\end{subequations}
Equation~\eqref{eq:phi0} is exact and holds generally for any electromagnetic field in Kerr.

\subsection{Horizon Regularity Conditions}
To proceed from Equation~\eqref{eq:phi0}, we need to consider the regularity conditions that relate the diverging parts of $E^i$ and $B^i$ at the horizon. For a general $F_{\mu\nu}$, we can find these regularity conditions by transforming to ingoing coordinates ($v,r,\theta,\tilde{\phi}$).
Performing this transformation, and
demanding that the Kerr-Schild field components remain finite, we find for a general $F_{\mu\nu}$ that the Boyer-Lindquist radial fields $E^r=E^r_{(\mathrm{KS})}$ and $B^r=B^r_{(\mathrm{KS})}$ must remain finite on the horizon. The Boyer-Lindquist angular fields may diverge, but must obey the regularity conditions:
\begin{equation}
\label{eq:regularity}
E_{-1}^\theta = -\frac{2Mr_+}{\Sigp}\sin\theta \, B_{-1}^\phi, \qquad
B^\theta_{-1} = \hphantom{+}\frac{2Mr_+}{\Sigp}\sin\theta \, E_{-1}^\phi,
\end{equation}
where the finite residues of the angular fields are $E^i_{-1}=\lim_{\D\rightarrow 0}\left(\D E^i\right)$ and $B^i_{-1}=\lim_{\D\rightarrow 0}\left(\D B^i\right)$ for $i\in\{\theta,\phi\}$.
These general regularity conditions are the origin of the Znajek condition (Equation~\eqref{eq:znajek1}); for stationary, axisymmetric, degenerate fields, the toroidal $E^\phi=0$ everywhere and $E^\theta=\frac{\Pi\sin\theta}{\Delta\Sigma}(\omega-\OmegaF)B^r$. Plugging these relations into Equation~\eqref{eq:regularity} shows that for stationary, axisymmetric degenerate fields $B^\theta=B^\theta_{(\mathrm{KS})}$ remains finite at the horizon and $B^\phi$ obeys Equation~\eqref{eq:znajek1}.

\subsection{$\Phi_0$ and $\mathcal{Z}_0$ on the Horizon}
Taking $\Phi_0$ in Equation~\eqref{eq:phi0} to the horizon while applying the regularity conditions (Equation~\eqref{eq:regularity}), we find that
\begin{equation}
\label{eq:phi0reg}
\Phi_0|_{r=r_+} = \frac{(r_+-ia\cos\theta)\sin\theta}{2\sqrt{2}Mr_+}\left(B_{-1}^\phi-i E_{-1}^\phi\right),
\end{equation}
which holds at the horizon for any electromagnetic field in Kerr. Multiplying by $p^m$ from Equation~\eqref{eq:pm} we have that, in general
\begin{equation}
\label{eq:pmphi0}
\left[p^m\Phi_0\right]_{r=r_+} = \frac{E\sin\theta\left(\beta_{\rm s}-i\nu_{\rm s}\right)}{4Mr_+}\left(B_{-1}^\phi-i E_{-1}^\phi\right).
\end{equation}
In deriving Equation~\eqref{eq:pmphi0}, we made no assumption that the field was even degenerate. However, if we specialize to the degenerate case, we can then use the general relationship of Equation~\eqref{eq:pmphi0} to calculate the observed EVPA from synchrotron emission from degenerate fields (Equation~\eqref{eq:Z0nullgen}) as
\begin{align}
\label{eq:generalZ0}
\mathcal{Z}_0 = \frac{\beta_{\rm s} +q \nu_{\rm s}}{\nu_{\rm s} - q\beta_{\rm s}},
\end{align}
where we define the ratio of the toroidal electric and magnetic field values at the horizon as
\begin{equation}
\label{eq:qdef}
q \equiv -\frac{E_{-1}^\phi}{B_{-1}^\phi} = +\left[\frac{\hphantom{\star}F_{t\phi}}{\star F_{t\phi}}\right]_{r=r_+}.
\end{equation}

\subsection{Stationary and Axisymmetric Case}
\label{app:npderivationstataxi}
Thus, we find that there is a universal horizon value of the EVPA from synchrotron radiation in degenerate fields so long as $q=0$, that is, when the toroidal electric field vanishes compared to the toroidal magnetic field at the horizon. The condition $q=0$ is met when the assumed degenerate field is axisymmetric and time-stationary, as is immediately obvious from the definition of $F_{\mu\nu}$ in terms of the vector potential $A_\mu$:
\begin{equation}
F_{t\phi} = -\Delta\sin^2\theta E^\phi = \partial_t A_\phi - \partial_\phi A_t = 0.
\end{equation}
Thus, in the stationary, axisymmetric case, the polarization becomes entirely determined by the diverging toroidal magnetic field at the horizon, which fixes the Penrose-Walker constant uniquely as
\begin{equation}
\label{eq:Z0v2}
\mathcal{Z}_0 = \frac{\beta_{\rm s}}{\nu_{\rm s}}.
\end{equation}

Our derivation of Equation~\eqref{eq:Z0v2} required assuming $\Phi_0\neq0$ at the horizon, which fails to hold if the toroidal field exactly vanishes, $B^\phi_{-1}=0$; for instance, the toroidal field vanishes in stationary, axisymmetric, degenerate magnetospheres when $\OmegaF=\OmegaH$ exactly by the Znajek condition (Equation~\eqref{eq:znajek1}). In this case, $q$ (Equation~\eqref{eq:qdef}) is indeterminate and the value of $\mathcal{Z}$ at the horizon must be obtained by expanding $\mathcal{A}$ and $\mathcal{B}$ to subleading order in $\Delta$. We find that in this edge case $\mathcal{Z}$ at the horizon does not take on the universal form and depends on the structure of the magnetic field, even if the field is axisymmetric, degenerate, and time-stationary. However, our derivation holds for all $B^\phi_{-1}\neq 0$, even when the value of $|B^\phi_{-1}|$ is arbitrarily small.

\section{Derivation of the Horizon Value $\mathcal{Z}_0$ in the ZAMO frame}
\label{app:zamoderivation}
In this Appendix, we re-derive the horizon value of $\mathcal{Z}_0$ (Equation~\eqref{eq:Z0general}) by working entirely in the Boyer-Lindquist normal observer, or zero-angular momentum (``ZAMO'') frame. We show that the universal polarization signature emerges as a consequence of the ZAMO frame polarization vector $f^{(\hat a)}$ becoming completely aligned with the polar $(\hat{\theta})$ direction to leading order. As a result, we can intuitively view the universal polarization pattern on the horizon as corresponding to the lensed and parallel-transported lines of longitude of the black hole.

\subsection{Drift Velocity}
\label{app:driftframe}
For degenerate, magnetically dominated fields, the drift frame $u_\perp^\mu$ is the unique frame with vanishing electric field $e^\mu$ that minimizes the Lorentz factor relative to the ZAMO frame, $\gamma_\perp=-\eta_\mu u^\mu_\perp$:
\begin{equation}
u^\mu_\perp = \gamma_\perp \left(\eta^\mu - \frac{1}{B^2}\epsilon^{\mu\nu\alpha\beta}\eta_\nu E_\alpha B_\beta\right),
\end{equation}
where the Lorentz factor $\gamma_\perp = (1-E^2/B^2)^{-1/2}$ (the field must evidently be magnetically dominated for the drift velocity to be timelike). The drift frame magnetic field $b_\perp^\mu=-\Fd^{\mu\nu}u_\nu$ is purely spatial in the coordinate frame:
\begin{equation}
b_\perp^t = 0 \quad,\quad b_\perp^i=B^i/u^t.
\end{equation}

All other timelike frames with vanishing electric field are related to the drift frame by parallel boosts along the magnetic field direction; $u^\mu\propto u^\mu_\perp + x B^\mu$ (e.g. \citetalias{BHPI}, equation D142). We can derive the source polarization using the drift frame without loss of generality because the electromagnetic field tensor $F_{\mu\nu}$ is invariant under these parallel boosts (\citetalias{BHPI}, equation D145). Thus, by Equation~\eqref{eq:polvec2} the synchrotron vector $f^\mu$ computed in the drift frame will be the same in any frame $u^\mu$ with vanishing electric field. Note that for the remainder of this Appendix we drop the subscripts on the drift frame velocity $u^\mu_\perp$ and magnetic field vector $b^\mu_\perp$.

\subsection{Vectors in the ZAMO Frame}
The four-velocity of the ZAMO frame is $\eta_{\mu} = (-\alpha,0,0,0)$,
where $\alpha=\sqrt{\Delta\Sigma/\Pi}$ is the lapse in Boyer-Lindquist coordinates.
The inverse tetrad to the ZAMO frame orthonormal coordinates $(\hat{t}),(\hat{r}),(\hat{\theta}),(\hat{\phi})$, which are aligned with their corresponding directions in the Boyer-Lindquist lab frame, is
\begin{equation}
\label{eq:zamotetrad}
e^{(\hat t)}_\mu = -\eta_\mu, \qquad
e^{(\hat r)}_\mu = \sqrt{g_{rr}}\,\mathrm{d}r, \qquad
e^{(\hat \theta)}_\mu = \sqrt{g_{\theta\theta}}\,\mathrm{d}\theta, \qquad
e^{(\hat \phi)}_\mu =  \sqrt{g_{\phi\phi}}\left(\mathrm{d}\phi-\omega\mathrm{d}t\right).
\end{equation}

Using the tetrad above, the ZAMO frame components of the photon momentum (Equation~\eqref{eq:PhotonMomentum}) are
\begin{equation}
\label{eq:photonmomentumZAMO}
    p^{(\hat t)} = E \sqrt{\frac{\Pi}{\Delta \Sigma}}(1-\omega \lambda), \qquad
    p^{(\hat r)} = \pm_r E \sqrt{\frac{\mathcal{R}(r)}{\Sigma\Delta}}, \qquad
    p^{(\hat \theta)} = \pm_\theta E \sqrt{\frac{\Theta(\theta)}{\Sigma}}, \qquad
    p^{(\hat \phi)} = E\lambda\sqrt{\frac{\Sigma}{\Pi \sin^2\theta}}.
\end{equation}
The ZAMO-measured electric and magnetic fields are $\cE^{(\hat i)} = \alpha\sqrt{g_{ii}}\,E^i$ and $\cB^{(\hat i)} = \alpha\sqrt{g_{ii}}\,B^i$.\footnote{These pick up an extra factor of $\alpha$ from the lab frame $E^i,B^i$ because of the definition of the electric field (Equation~\eqref{eq:emubmu}) applied to the ZAMO frame; $\mathcal{E}^\mu=F^{\mu\nu}\eta_{\nu}, \mathcal{B}^\mu=-\Fd^{\mu\nu}\eta_{\nu}$.}
In the ZAMO orthonormal frame, the drift velocity is
\begin{equation}
\label{eq:u-drift}
    u^{(\hat t)} = \gamma \,,\quad
    \vec{u} = \frac{\gamma}{\cB^2}\,\vec\cE\times\vec\cB\,,
\end{equation}
where the Lorentz factor $\gamma = (1-\E^2/\B^2)^{-1/2}$.
The magnetic field in the drift frame is
\begin{equation}
\label{eq:b-drift}
    b^{(\hat t)} = 0\,,\quad
    \vec{b} = \vec\cB/\gamma.
\end{equation}
Using Equation~\eqref{eq:polvec}, the polarization vector in the ZAMO frame is then
\begin{equation}
\label{eq:f-ZAMO-temporal}
    f^{(\hat t)} = 0,\qquad
    \vec{f} = p^{(\hat t)}\vec\cE + (\vec{p}\times\vec\cB) - (\vec{p}\cdot\vec\cE)\frac{\vec{p}}{p^{(\hat t)}},
\end{equation}
where we use the definition $\vec b= \vec \cB/\gamma$, the drift frame identity $\vec{b}\times\vec{u} = \vec\cE$, the degenerate field condition $\vec\cE\cdot\vec\cB=0$, and we impose temporal gauge $f^{(\hat t)}=0$ in the ZAMO frame using Equation~\eqref{eq:Shift}.

\subsection{Near-Horizon Scalings}
\label{app:zamoscaling}
Using the general horizon regularity conditions (Equation~\eqref{eq:regularity}), we see that the radial components of the ZAMO $\mathcal{E}$ and $\mathcal{B}$ fields at the horizon are finite, but the angular components generally diverge as $\D^{-1/2}$.\footnote{Note that because the ZAMO-frame components carry a factor of the lapse $\alpha\sim\sqrt{\Delta}$, the near-horizon expansions in Appendix~\ref{app:zamoderivation} proceed in half-integer powers of $\Delta$ in contrast to the integer-power expansions of the Newman-Penrose (Appendix~\ref{app:npderivation}) and drift-frame (Appendix~\ref{app:driftderivation}) derivations.}
The diverging angular fields in the ZAMO frame are related by the general regularity conditions (Equation~\eqref{eq:regularity}):
\begin{equation}
\label{eq:regularityZ}
 \cE_{\circ}^{(\hat\theta)}=- \cB_{\circ}^{(\hat\phi)} \quad,\quad \cB_{\circ}^{(\hat\theta)}=\cE_{\circ}^{(\hat\phi)},
\end{equation}
where the residues of the leading $\D^{-1/2}$ divergences in the ZAMO frame angular magnetic fields are
\begin{equation}
    \cB_{\circ}^{(\hat \theta)}=\lim_{\D\to0}\left(\sqrt{\D}\cB^{(\hat \theta)}\right)=\frac{\Sigp}{2M\rp}\, B_{-1}^\theta \quad,\quad \cB_{\circ}^{(\hat \phi)}=\lim_{\D\to0}\left(\sqrt{\D}\cB^{(\hat \phi)}\right)=\sin\theta \,B_{-1}^\phi,
\end{equation}
and the electric field residues $\cE_{\circ}^{(\hat a)}$ are defined analogously.
The ZAMO frame electric and magnetic 3-vectors scale as
\begin{subequations}
\label{eq:ZAMOfields}
\begin{align}
\vec\cE &= \Delta^{-1/2} \left(0,-\cB_{\circ}^{(\hat\phi)},\cE_{\circ}^{(\hat\phi)}\right)+ \mathcal{O}(\D^0), \\
\vec\cB &= \Delta^{-1/2} \left(0,\hphantom{-}\cE_{\circ}^{(\hat\phi)},\cB_{\circ}^{(\hat\phi)}\right) + \mathcal{O}(\D^0).
\end{align}
\end{subequations}
The drift frame Lorentz factor also diverges as $\gamma \propto \Delta^{-1/2}$. In terms of the finite residue at the horizon $\gamma_{\circ}=\lim_{\D\to 0}\left(\sqrt{\D}\gamma\right)$, we find from Equation~\eqref{eq:u-drift} that the leading part of the drift frame velocity as seen by the ZAMO observer is radial and ingoing;
\begin{equation}
\label{eq:u-driftH}
    u^{(\hat a)} =\gamma_{\circ}\Delta^{-1/2} \left(1,-1,0,0\right)+ \mathcal{O}(\Delta^0).
\end{equation}
At leading order, the drift frame falls inward at the speed of light as viewed by the ZAMO observer. From Equations~\eqref{eq:b-drift},~\eqref{eq:u-driftH} and~\eqref{eq:ZAMOfields}, the drift frame magnetic field scales as
\begin{equation}
\label{eq:bZAMOscaling}
b^{(\hat a)} = \gamma^{-1}_{\circ}\left(0,0,\cE^{(\hat\phi)}_{\circ},\cB^{(\hat\phi)}_{\circ}\right) + \mathcal{O}(\Delta^{1/2}).
\end{equation}
The drift frame magnetic field as viewed from the ZAMO frame is finite on the horizon and the angular parts dominate over the radial part. When $q=0$, only the toroidal component $b^{(\hat\phi)}$ survives to leading order.

The photon momentum as seen by the ZAMO observer (Equation~\eqref{eq:photonmomentumZAMO}) becomes radial and outgoing at the horizon:
\begin{equation}
\label{eq:pZAMOscaling}
p^{(\hat a)} = \mathcal{D}\Delta^{-1/2}\Bigl(1,1,0,0\Bigr) + \Bigl(0,0,p^{(\hat\theta)}_0,p^{(\hat\phi)}_0 \Bigr) + \mathcal{O}(\Delta^{1/2}).
\end{equation}
The leading order coefficient in Equation~\eqref{eq:pZAMOscaling} is $\mathcal{D}=E\Pp/\sqrt{\Sigp}$, where $\Pp=\sqrt{\mathcal{R}}|_{\rp}=2M\rp-a\lambda$. The subleading angular parts are $p^{(\hat\theta)}_0 = E\beta_{\rm s}/\sqrt{\Sigp}$ and $p^{(\hat\phi)}_0 = E\lambda\sqrt{\Sigp}/(2Mr_+\sin\theta)$.

Using these horizon scalings in Equation~\eqref{eq:f-ZAMO-temporal}, we see that for general finite $q$, the angular components of polarization vector in temporal gauge dominate at $O(\D^{-1})$:
\begin{equation}
\label{eq:fZAMOscaling}
f^{(\hat a)} = 2\mathcal{D}\,\Delta^{-1}\Bigl(0,0,-\cB_{\circ}^{(\hat \phi)},\cE_{\circ}^{(\hat \phi)}\Bigr) + \D^{-1/2}\Bigl(0,\mathcal{F}^{(\hat r)},0,0\Bigr) + \mathcal{O}(\Delta^0),
\end{equation}
where the subleading term is $\mathcal{F}^{(\hat r)}=2\big(p_0^{(\hat \theta)}\cB_{\circ}^{(\hat \phi)}-p_0^{(\hat\phi)}\cE_{\circ}^{(\hat \phi)}\big)$.
When the degenerate field is stationary and axisymmetric ($E^{\phi}=0$) the predominant part of the horizon polarization vector is in the polar $\hat{\theta}$ direction, with a subleading correction in the radial direction $f^{(\hat r)}/f^{(\hat\theta)}\sim \D^{1/2}$.

We can physically interpret the fact that the ZAMO-frame horizon polarization for synchrotron radiation from stationary, axisymmetric, degenerate magnetospheres is predominantly polar as a necessary result of these three horizon limits: (1) the drift frame velocity is predominantly ingoing (Equation~\eqref{eq:u-driftH}; timelike emitters at the horizon must be directed to the BH interior); (2) the divergent photon momentum is predominantly outgoing (Equation~\eqref{eq:pZAMOscaling}; the closing of the outgoing photon light cone); and (3) the magnetic field becomes dominated by the toroidal component (Equation~\eqref{eq:bZAMOscaling}). Thus, to leading order, the only mutually perpendicular direction that the polarization can take at the horizon is aligned with the polar direction. In a general degenerate magnetosphere when $q\neq0$, the emitter frame magnetic field also has a component in the $\hat{\theta}$ direction, and the universal behavior forcing the emitted polarization to be predominantly polar is lost.

\subsection{Value of $\mathcal{Z}_0$}
In temporal gauge, the $\mathcal{A}$ and $\mathcal{B}$ components of the Penrose-Walker constant are (see e.g. \citetalias{BHPI}, equation G13):
\begin{subequations}
\label{eq:ABZAMO}
\begin{align}
    \cA &= \frac{1}{\sqrt{\Pii}}\bigl[
        ((r^2+a^2) p^{(\hat t)} - a\sin\theta\sqrt{\D}\,p^{(\hat\phi)})\, f^{(\hat r)}
        + a\sin\theta\sqrt{\D}\,p^{(\hat r)}\, f^{(\hat\phi)}
    \bigr],\label{eq:A-tg}\\
    \cB &= \frac{1}{\sqrt{\Pii}}\bigl[
        (-a\sin\theta\sqrt{\D}\,p^{(\hat t)} + (r^2+a^2)p^{(\hat\phi)})\, f^{(\hat\theta)}
        - (r^2+a^2)p^{(\hat\theta)}\, f^{(\hat\phi)}
    \bigr].\label{eq:B-tg}
\end{align}
\end{subequations}
Using the scalings in Appendix~\ref{app:zamoscaling}, we see that for general finite $q$ every term in Equation~\eqref{eq:A-tg} and Equation~\eqref{eq:B-tg} contributes at $\mathcal{O}(\Delta^{-1})$ except the $p^{(\hat \phi)}f^{(\hat r)}$ term, which is $\mathcal{O}(\Delta^0)$. While in the ZAMO frame the photon momentum becomes outgoing to leading order and the polarization becomes angular, in deriving the polarization seen at infinity we must include the subleading parts $p^{(\hat\phi)},p^{(\hat\theta)}$ (in $\mathcal{B}$) and $f^{(\hat r)}$ (in $\mathcal{A}$).

Assuming that $\cB_{\circ}^{(\hat \phi)}$ is nonzero and
substituting the leading and sub-leading order expressions from Equation~\eqref{eq:fZAMOscaling} and Equation~\eqref{eq:pZAMOscaling} into Equation~\eqref{eq:ABZAMO} gives:
\begin{equation}
    \lim_{\D\to0}\D\cA = 2\mathcal{D} \cB_{\circ}^{(\hat \phi)}\left[p^{(\hat\theta)}_0 + q\left(p^{(\hat\phi)}_0-\frac{a\mathcal{D}\sin\theta}{2Mr_+}\right)\right],\quad
    \lim_{\D\to0}{\D\cB} = 2\mathcal{D} \cB_{\circ}^{(\hat \phi)}\left[q\,p^{(\hat\theta)}_0-\left(p_0^{(\hat\phi)}-\frac{a\mathcal{D}\sin\theta}{2Mr_+}\right)\right].
\end{equation}
Inserting $\mathcal{D}=E\Pp/\sqrt{\Sigp}$, $p^{(\hat\theta)}_0 = E\beta_{\rm s}/\sqrt{\Sigp}$, and $p^{(\hat\phi)}_0 = E\lambda\sqrt{\Sigp}/(2Mr_+\sin\theta)$, we find
\begin{equation}
    \lim_{\D\to0}\D\cA = \frac{2E\,\mathcal{D} \cB_{\circ}^{(\hat \phi)}}{\sqrt{\Sigp}}\left(\beta_{\rm s}+q\nu_{\rm s}\right),\quad
    \lim_{\D\to0}{\D\cB} = \frac{2E\,\mathcal{D}\cB_{\circ}^{(\hat \phi)}}{\sqrt{\Sigp}}\left(q\beta_{\rm s}-\nu_{\rm s}\right).
\end{equation}
Thus, we find from the ZAMO frame derivation that the ratio $\mathcal{Z}_0=-\lim_{\D\to0}\mathcal{A}/\mathcal{B}$ reduces to
\begin{equation}
\label{eq:Z0v3}
    \mathcal{Z}_0 = \frac{\beta_{\rm s} + q\nu_{\rm s}}{\nu_{\rm s} - q\beta_{\rm s}}\,,
\end{equation}
in agreement with the above result, Equation~\eqref{eq:generalZ0}.

\section{Derivation of the Horizon Value $\mathcal{Z}_0$ in the drift frame}
\label{app:driftderivation}
In this Appendix, we re-derive the horizon value of $\mathcal{Z}_0$ (Equation~\eqref{eq:Z0general}) by working in the plasma drift frame (Appendix~\ref{app:driftframe}).
Here we show that, as in the ZAMO frame derivation in Appendix~\ref{app:zamoderivation}, the universal polarization signature emerges as a consequence of the drift frame polarization vector becoming aligned with the rest-frame analogue of the global $\hat{\theta}$ direction. When compared to the ZAMO frame derivation above, the tetrad transformation to the drift frame is more complicated; however, this version of the derivation has the advantage that the drift frame is timelike at the horizon; all divergences in this frame at the horizon can then be traced back to the diverging energy $E'$ of outgoing photons at the horizon.

\subsection{Drift Frame Tetrad}
We define a tetrad for the drift frame coordinates $((\hat0),(\hat1),(\hat2),(\hat3))$ that has the $(\hat1)$ direction aligned with the local photon momentum and the $(\hat 2)$ direction aligned with the projected azimuthal Killing vector $\xi_{(\phi)}=\partial_\phi$:\footnote{Note the change in the order of the drift frame basis directions from the associated coordinate directions. The $(\hat 2)$ direction is associated with the projected coordinate $\hat{\phi}$ direction; when the photon momentum becomes radial the $(\hat 3)$ direction is associated with the -$(\hat \theta)$ direction.}
\begin{subequations}
\label{eq:drifttet}
\begin{align}
\label{eq:drifttet0}
e_{(\hat0)}^\mu &= u^\mu, \\
\label{eq:drifttet1}
e_{(\hat1)}^\mu &= \frac{p^\mu}{E'} - u^\mu, \\
\label{eq:drifttet2}
e_{(\hat2)}^\mu &= \frac{1}{N}\,S^\mu{}_\nu \ \xi_{(\phi)}^\nu,\\
\label{eq:drifttet3}
e_{(\hat3)}^\mu &= -\,\epsilon^\mu{}_{\nu\alpha\beta}\,u^\nu e_{(\hat1)}^\alpha e_{(\hat2)}^\beta,
\end{align}
\end{subequations}
where the local photon energy $E'=-p_\mu u^\mu$ and the screen projector $S^{\mu\nu}$ is
\begin{equation}
S^{\mu\nu} = g^{\mu\nu} + u^\mu u^\nu - e_{(\hat1)}^\mu e_{(\hat1)}^\nu.
\end{equation}
Because $S$ is idempotent, the normalizing factor in Equation~\eqref{eq:drifttet2} is $N^2 = S_{\mu\nu} \, \xi_{(\phi)}^\mu \xi_{(\phi)}^\nu$.

\subsection{Vectors in the Drift Frame}
By construction, the drift velocity is aligned along the $(\hat 0)$ direction and the spatial part of the photon momentum is aligned along the $(\hat 1)$ direction:
\begin{align}
\label{eq:udrift}
u^{(\hat a)} &= (1,\,0,\,0,\,0), \\
\label{eq:kdrift}
p^{(\hat a)} &= E'\,(1,\,1,\,0,\,0).
\end{align}
The photon energy $E'$ of the outgoing photon measured in the infalling drift frame also diverges at the horizon as
\begin{equation}
E' = \frac{2EU\Pp}{\Delta} + E'_0 + \mathcal{O}(\Delta),
\end{equation}
where $U\equiv-u^r(r=\rp)$ is the finite ingoing radial velocity at the horizon.

To compute the magnetic field vector at the horizon, we
recall from Appendix~\ref{app:driftframe} that the drift frame field is related to the coordinate frame field by $b^t=u\cdot B=0$, $b^i=B^i/u^t$. Regularity at the horizon (Equation~\eqref{eq:regularity}) implies the coordinate frame radial field $B^r$ is $\mathcal{O}(1)$ while in general $B^\theta,B^\phi$ diverge as $\mathcal{O}(\Delta^{-1})$.
As a result, the drift frame radial field vanishes in Boyer-Lindquist coordinates, $b^r\propto\Delta$, while $b^\theta$ and $b^\phi$ are $\mathcal{O}(1)$.
Transforming $b^\mu$ to the drift frame basis by expanding the tetrad Equation~\eqref{eq:drifttet} order-by-order in $\Delta$, we find that similarly $b^{(\hat 0)}=0$ by construction while $b^{(\hat 1)}\propto \Delta$. Only the screen components of the drift frame magnetic field survive and are finite at leading order:
\begin{align}
\label{eq:bdrift}
b^{(\hat a)} &=\frac{1}{U\sqrt{\Sigp}}\, \left(0,\ \ 0,\ \ \cB^{(\hat \phi)}_{\circ},\
-\cE^{(\hat \phi)}_{\circ}\right) + \mathcal{O}(\Delta),
\end{align}
where we have used the horizon regularity condition for the ZAMO-frame fields $\cB^{(\hat \theta)}_{\circ}=\cE^{(\hat \phi)}_{\circ}$ (Equation~\eqref{eq:regularityZ}). As in the ZAMO frame derivation in Appendix~\ref{app:zamoderivation}, we see from Equation~\eqref{eq:bdrift} that when the field is axisymmetric and time-stationary, the drift frame magnetic field becomes aligned with the local azimuthal direction at the horizon.

The leading order polarization direction $\vec{f}=\vec{p}\times\vec{b}$ at the horizon is then simply
\begin{equation}
f^{(\hat a)} = \frac{E'}{U\sqrt{\Sigp}}\,\left(0,\ 0,\ \cE^{(\hat \phi)}_{\circ},\ \cB^{(\hat \phi)}_{\circ}\right) + \mathcal{O}(\Delta^0).
\label{eq:fproj}
\end{equation}
Since $E'\propto \D^{-1}$ diverges at the horizon, the polarization vector also diverges in the local frame.
\subsection{Penrose-Walker Constant}
The Penrose-Walker constant in the drift frame is
\begin{align}
\kappa &= \left(\star J_{(\hat a)(\hat b)} + i\,J_{(\hat a)(\hat b)}\right)\,p^{(\hat a)} f^{(\hat b)}.
\end{align}
Since $p^{(\hat a)}=E'(1,1,0,0)$ by construction and $f^{(\hat a)}=(0,0,f^{(\hat 2)},f^{(\hat 3)})$ to leading order in the drift frame at the horizon, the imaginary part $\kappa_2$ at the horizon is
\begin{equation}
    \kappa_2 = E'\left(P_2 f^{(\hat 2)} + P_3 f^{(\hat 3)}\right),
\end{equation}
where $P_{(\hat a)}= J_{(\hat 0)(\hat a)} + J_{(\hat 1)(\hat a)}$. By the duality of $\star J$ and $J$, we find that $\kappa_1/E' = -P_3 f^{(\hat 2)} + P_2 f^{(\hat 3)}$, so in combination
\begin{equation}
\label{eq:kappafactored}
    \kappa = E'\left(P_2 + i P_3\right)\big(f^{(\hat 3)} + i f^{(\hat 2)}\big).
\end{equation}
What remains is to calculate $P_2$ and $P_3$. Using Equation~\eqref{eq:KYBL} and carefully expanding the tetrad Equation~\eqref{eq:drifttet} order-by-order in $\Delta$, we find that
\begin{equation}
P_2 + iP_3 = \frac{\Delta}{2U\Pp\sqrt{\Sigp}}\,(\rp - i a\cos\theta)\,(\beta_{\rm s} + i\nu_{\rm s}) + \mathcal{O}(\Delta^2).
\label{eq:P23complex}
\end{equation}
Substituting Equation~\eqref{eq:fproj} and
Equation~\eqref{eq:P23complex} into Equation~\eqref{eq:kappafactored} and using $E' = 2EU\Pp/\Delta$ to leading order, we obtain (assuming $\cB_{\circ}^{(\hat \phi)}$ is nonzero):
\begin{equation}
\kappa = \frac{2E^2\Pp}{\Delta\Sigp}\,\cB^{(\hat\phi)}_{\circ}\left(\rp - i a\cos\theta\right)\,\left(\beta_{\rm s} + i\nu_{\rm s}\right)\,(1 -iq) + \mathcal{O}(\D^0),
\label{eq:kappafull}
\end{equation}
where $q=-\cE^{(\hat \phi)}_{\circ}/\cB^{(\hat \phi)}_{\circ}$. From Equation~\eqref{eq:kappafull}, using the decomposition $\kappa=(r-ia\cos\theta)(\mathcal{A}-i\mathcal{B})$, we can read off the ratio $\mathcal{Z}_0=-\mathcal{A}/\mathcal{B}$ at the horizon:
\begin{equation}
    \mathcal{Z}_0 = \frac{\beta_{\rm s} + q\nu_{\rm s}}{\nu_{\rm s} - q\beta_{\rm s}},
\end{equation}
which agrees with our derivations in the Newman-Penrose formalism (Appendix~\ref{app:npderivation}) and in the ZAMO frame (Appendix~\ref{app:zamoderivation}).

\bibliography{bhp3}
\bibliographystyle{aasjournalv7.1}

\end{document}